\documentclass[article]{IEEEtran}

\usepackage{cite}
\usepackage{graphicx}
\usepackage{rotating}

\begin{document}
\title{Impact of Report Message Scheduling (RMS) in 1G/10G EPON and
  GPON (Extended Version)}

\author{Anu~Mercian, Michael P.~McGarry, and Martin~Reisslein
\thanks{Technical Report, School of Electrical, Computer, and Energy Eng.,
Arizona State University, December 2013. This extended technical
report accompanies~\cite{MeMR14}.
 Please direct correspondence to M.~Reisslein.}
\thanks{A.~Mercian and M.~Reisslein
are with the School of Electrical, Computer, and Energy Engineering,
Arizona State University, Tempe, Arizona 85287-5706,
E-mail: \{amercian, reisslein\}@asu.edu,
Phone: (480)965-8593, Fax: (480)965-8325.}
\thanks{M.~McGarry is with the Dept.\ of Electrical and Computer Eng.,
University of Texas at El Paso,
500 W University Ave, El Paso, TX 79968,
Email: mpmcgarry@utep.edu,
Phone: (915)747-6955, Fax: (915)747-7871.}
}

\maketitle

\begin{abstract}
A wide array of dynamic bandwidth allocation (DBA) mechanisms have
recently been proposed for improving bandwidth utilization and
reducing idle times and packets delays in passive optical networks
(PONs). The DBA evaluation studies commonly assumed that the report
message for communicating the bandwidth demands of the distributed
optical network units (ONUs) to the central optical line terminal
(OLT) is scheduled for the end of an ONU's upstream transmission,
after the ONU's payload data transmissions. In this article, we
conduct a detailed investigation of the impact of the report message
scheduling (RMS), either at the beginning (i.e., before the pay load
data) or the end of an ONU upstream transmission on PON performance.
We analytically characterize the reduction in channel idle time with
reporting at the beginning of an upstream transmission compared to
reporting at the end. Our extensive simulation experiments consider
both the Ethernet Passive Optical Networking (EPON) standard and the
Gigabit PON (GPON) standard. We find that for DBAs with offline
sizing and scheduling of ONU upstream transmission grants at the end
of a polling cycle, which processes requests from all ONUs,
reporting at the beginning gives substantial reductions of mean
packet delay at high loads. For high-performing DBAs with online
grant sizing and scheduling, which immediately processes individual
ONU requests, or interleaving of ONUs groups, both reporting at the
beginning or end give essentially the same average packet delays.
\end{abstract}


\section{Introduction}
\label{intro:sec}
\begin{figure}[t!]
\centering
\includegraphics[scale=0.48]{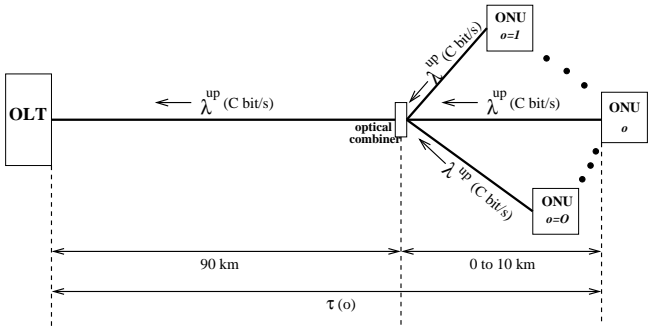}
\caption{Illustration of upstream transmission direction
from distributed ONUs $o,\ o = 1, 2, \ldots, O$,
to a central OLT in the PON structure.
The $O$ ONUs share a single wavelength channel with bit rate
$C$ bit/s for their upstream transmissions and have one-way
propagation delay $\tau(o)$ to the OLT.}
\label{arch:fig}
\end{figure}

Passive optical networks (PONs) have emerged over the past decade
as a highly promising access network technology for
connecting individual distributed optical network units (ONUs)
at distributed subscriber premises to a central
optical line terminal
(OLT), see Fig.~\ref{arch:fig},~\cite{AuSMR10,ChAKLS10,ChWCF11,Dixit2013,EfE09,KeLE10,KrDR12,MaMCW13,SaS13,WeAMR13}.
Recent advances in the underlying photonic
and physical layer communications
technologies and commensurate standardization efforts
have paved the way for PONs operating at a channel
bandwidth of 10~Gbps
(compared to the 1~Gbps bandwidth considered in early PON
development), cf.~IEEE 802.3av~\cite{TanAH2010} and G.987~\cite{XGPON}.
Also, long-reach PONs operating up to distances of 100~km
between the distributed ONUs and the central OLT have
emerged~\cite{AnBT12, Mou02, Mou05, LiG13, ScBLP11, ShND13, SiSS12, SKM0110}.
Operating at high bandwidth over long distances,
i.e., with a high bandwidth-delay product, poses significant
challenges for coordinating the upstream transmissions of the
distributed ONUs so as to avoid collisions on the
shared upstream (from ONUs to OLT)
channel.
As reviewed in Section~\ref{lit:sec},
a wide array of dynamic bandwidth allocation (DBA) mechanisms
have been developed to solve this medium access control problem
on the upstream channel for bursty ONU packet traffic.

The DBA mechanisms operate commonly within the context of the
standardized signaling mechanisms for PONs which are based on
a cyclical report-grant polling structure,
which is illustrated in Fig~\ref{fig:cycle_illu}.
More specifically, ONUs signal their queue depths, i.e.,
current bandwidth demands, with a control (report) message to the OLT.
The OLT then sets the sizes (lengths) of the upstream transmission
windows (grants) for the individual ONUs and signals the length
and starting time (schedule) of each transmission window
to the individual ONUs through grant messages,
which are represented by the downward arrows in Fig~\ref{fig:cycle_illu}.
In particular, the Ethernet PON (EPON) employs the
Report and Gate messages of the Multi-point Control Protocol
(MPCP) according to the IEEE 802.3ah or 802.3av standards.
The Gigabit PON (GPON) employs dynamic bandwidth reports
upstream (DBRu) for signaling the queue depths and
Bandwidth Maps (BWMaps) for signaling the upstream transmission windows
following the G.984 or G.987 standards~\cite{HoTr12}.

The report message from an ONU is typically lumped together with
the upstream payload data transmission so as to avoid extra
guard times for the short report message.
While the EPON standard leaves the position of the report message
within an ONU's upstream transmission open, the vast majority of
EPON studies have assumed that the report message is
positioned at the very end of an ONU's upstream transmission,
after the ONU's payload data transmissions.
This ``reporting at the end'' allows the ONU to signal the most
up-to-date queue depth, at ideally the time instant of the
end of the payload transmission, to the OLT.
On the other hand, the GPON standard specifies that the report message
be included at the beginning of the upstream transmission, i.e.,
to precede the payload data~\cite{XGPON}.
This ``reporting at the beginning'' has the advantage that
the OLT receives the report message earlier (i.e., before the
ONUs payload data) and can already size and schedule the transmission windows
for the next cycle. On the downside, the report at the beginning does not
contain the packets that were newly generated during the
ONU's payload transmission.

To the best of our knowledge the impact of the report message
scheduling at the beginning or end of an ONU's upstream transmission
has not yet been investigated in detail.
In this article we examine this open research question
in the context of both EPONs and GPONs operating at either
1~Gbps or 10~Gbps channel bandwidth for state-of-the-art DBA mechanisms.
We analyze the channel idle time with reporting at the beginning or end
for the different DBA mechanisms.
We show that reporting at the beginning can reduce the
channel idle time that precedes the arrival of an ONU
upstream transmission at the OLT by up to the
transmission time of an ONU's payload compared to reporting at the end.
We conduct extensive simulations to evaluate the average packet
delay.
We find that reporting at the beginning significantly reduces
the packet delay for DBA mechanisms that accumulate all reports
from a cycle for offline transmission window sizing and scheduling.
In contrast, DBA mechanisms that size and schedule transmission
windows online or employ interleaving strategies for the cyclic
polling processes, perform equally well for both reporting
at the beginning or end.

This article is structured as follows.
Section~\ref{lit:sec} discusses the background and related work.
In Section~\ref{con:sec}, we analyze the channel idle time
for reporting at the beginning and identify the reduction in
channel idle time compared to reporting at the end.
In Section~\ref{perf:sec}, we present extensive
simulation results comparing reporting in the
beginning and end for 1~Gbps and 10~Gbps EPON and GPON.
In Section~\ref{concl:sec}, we summarize our observations
and outline future research directions.

\section{Background and Related Work}
\label{lit:sec}
\subsection{Dynamic Bandwidth Allocation (DBA)}
Efficient control of the access by the distributed ONUs to the
shared upstream channel so as to serve bursty traffic with low delay
while avoiding collisions is one of the key challenges in
operating a PON~\cite{GuPT12}.
Several dynamic bandwidth allocation (DBA) approaches have
been developed for this channel access problem.
A primary classification criterion for DBA mechanisms is the
number of polling threads employed per ONU.
Single-thread polling
(STP)~\cite{AuSR11,DeSG10,KMP0202,LiRo13,LA0205,RaR11,JiMFD12,KT0809,QiXZ13}
employs one polling thread per ONU, while multi-thread polling
(MTP)~\cite{BuAT13,Mou05,Bur03,MeMR13,Muk01}
employs multiple polling threads.
The polling threads may operate in offline fashion, i.e.,
collect report messages from all ONUs before sizing and scheduling the
upstream transmission windows, or make these decisions in online
fashion after the receipt of each individual report
message~\cite{ZhMo09}.

The vast majority of the existing studies on
DBA in PONs has considered reporting at the end of the upstream transmission.
Reporting at the beginning
has only briefly been considered for STP with elementary
gated grant sizing in~\cite{AuSH08} and for MTP in~\cite{DixDLCPD2011}.
Also, the channel idle time has so far only been analyzed
for reporting at the end of an upstream transmission
in~\cite{Mar01,MeMR13}.
The present study provides the first analysis of the channel idle time
for reporting at the beginning of an ONU upstream transmission
as well as a detailed examination of
the impact of the report message scheduling at beginning vs. end of
an upstream transmission on the channel idle time and packet delay
for a wide range of DBA mechanisms.

\subsection{PON Standards}
PONs with 1~Gbps channel bandwidth were standardized a decade ago as
EPON in IEEE~802.3ah~\cite{Kramer05}
and as GPON in ITU-T~G.984~\cite{GPON}. On the other hand,
corresponding standards for 10~Gbps channel bandwidth were established
only recently as 10G-EPON in IEEE 802.3av~\cite{TanAH2010}
and as XG-PON in ITU-T G.987~\cite{XGPON}.
As a result, DBA mechanisms for 10~Gbps EPON or XG-PON have
yet to receive significant attention from the research
community. Several comparisons of the physical layer and link layer
overhead differences among the various 1~Gbps and 10~Gbps standards have
appeared in the literature~\cite{RoyKHS11,BegHR2011,HajSM2007,Tal01}.

Some early investigations of DBA mechanisms
for the 10~Gbps standards have been reported
in~\cite{SCAW09,IK1209,GutGS2011,Han2013,HanYL2013}.
The impact of the polling cycle time in
single-thread polling with limited grant sizing on various
performance measures, e.g., packet delay and jitter, for each of the
1~Gbps and 10~Gbps standardized PON variants was studied in~\cite{SCAW09}.
Mechanisms to increase TCP throughput for 10~G-EPON
were studied in~\cite{IK1209}.
A modification to an existing DBA algorithm to support a mixed
network of both 1~Gbps and 10~Gbps EPON ONUs was proposed in~\cite{GutGS2011}.
Efficient utilization of unused bandwidth for
XG-PON was investigated in~\cite{Han2013,HanYL2013}.
The work presented in this article augments
this relatively small body of literature by investigating the impact
of the reporting position on performance measures for each of the
1~Gbps and 10~Gbps standardized PON variants.

\section{Analysis of Channel Idle Time}
\label{con:sec}
\subsection{PON Model}
In this section we analyze the idle time on the upstream channel of a PON
before each upstream transmission.
We consider a PON model with a total number of $O$ ONUs,
whereby ONU $o,\ o = 1, 2, \ldots, O$, has a one-way
propagation delay of $\tau(o)$ [s] to the OLT.
The $O$ ONUs share the upstream wavelength channel with
bit rate $C$ [bit/s], as illustrated in Fig~\ref{arch:fig}.
Polling-based medium access control with
report-grant cycles is employed to avoid collisions on
the shared upstream wavelength channel.
We denote $n$ for the polling cycle index and $t_g$ for the
guard time, i.e., the minimum required spacing between
successive upstream transmissions from different ONUs.
Moreover, we denote $t_R$ and $t_G$ for the transmission times
of a report and grant message, respectively, as
summarized in Table~\ref{not:tab}.

For DBA mechanisms employing multiple polling
threads~\cite{BuAT13,Mou05,Bur03,MeMR13,Muk01}, we denote $\Theta$
for the total number of threads, with $\theta,\ \theta = 1, 2,
\ldots, \Theta$, denoting the thread index.
The $\Theta$ threads operate in parallel, giving each ONU
$\Theta$ opportunities to report the queue depth and transmit
upstream payload data in a polling cycle.
Note that $\Theta = 1$
corresponds to single-thread polling.
We omit the thread index $\theta$ from the model notations for single-thread
polling.
We denote $Z$ for the maximum
cycle duration in terms of the sum (aggregation) of the upstream
transmissions of all $O$ ONUs (and all $\Theta$ threads) of a given
cycle $n$. A particular grant scheduling policy may arrange the
upstream transmission windows of the $O$ ONUs of a given thread
$\theta$ in a particular order. We use the index $j,\ j = 1, 2,
\ldots, O$ to denote the ordering of the ONU upstream transmissions
(of a given thread $\theta$) in a given cycle $n$.

\begin{table}[t]
\caption{PON modeling parameters}
\label{not:tab}
\centering
\begin{tabular}{|l|l|}
\hline
Parameter & Meaning \\ [0.5ex]
\hline
\multicolumn{2}{|c|}{Network and polling structure} \\
$O$     &   Total number of ONUs, indexed $o = 1, 2, \ldots, O$ \\
$\tau(o)$   &   One-way propagation delay from OLT to ONU $o$ [s]\\
$C$  & Upstream bandwidth [bit/s] \\
$Z$             & Maximum cycle duration, i.e., max. aggregate duration \\
                &   of upstream transmission windows of all $O$ ONUs (and \\
           & all $\Theta$ threads) in a cycle [s] \\
$t_R$  & Transmission time of a report message [s] \\
$t_g$ & Guard time [s] \\
$t_G$ & Gate transmission time [s] \\
\hline
\multicolumn{2}{|c|}{Cycle, thread, and upstream transmission indices} \\
$n$     &   Polling cycle index \\
$\Theta$        &   Total number of threads \\
$\theta$        &   Thread index, $\theta = 1, \ldots, \Theta$ \\
$j$ &  ONU index ordered by upstream transm.\ position for a \\
     & given thread $\theta$ in  a given cycle $n$, i.e., ONU $j$ has $j$th \\
        & upstream transmission grant of thread $\theta$ in cycle $n$ \\
\hline
\multicolumn{2}{|c|}{Upstream transmission window (grant) scheduling} \\
\multicolumn{2}{|c|}{[all parameters are in units of seconds]} \\
$\gamma(n,\theta, j)$ & Time instant when OLT makes scheduling decision for \\
  & transmission window of $j$th ONU of thread $\theta$ in cycle $n$ \\
$T(n, \theta, j)$ & Gate signaling delay: Time duration from instant of OLT  \\
                &  scheduling decision to end of the GATE transm.\ for $j$th\\
  & ONU of thread $\theta$ in cycle $n$ plus round-trip prop.\ delay \\
$\alpha(n,\theta, j)$ & Time instant when upstream transmission of $j$th ONU \\
 & of thread $\theta$ in cycle $n$ starts to arrive at OLT\\
$\beta(n,\theta, j)$ &  Time instant when end of upstream transm.
of $j$th ONU \\
   & of thread $\theta$ in cycle $n$ arrives at OLT. \\
$\Omega(n, \theta, j)$ & Time instant of end of upstream transm.\
  preceding arrival \\
  & of upstream transm. of $j$th ONU of thread $\theta$ in cycle $n$\\
$I(n, \theta, j)$ & Duration of channel idle time preceding the
   arrival of up-  \\
  & stream transm.\ of $j$th ONU of thread $\theta$ in cycle $n$ at OLT\\
$G_{\max} $  & = $\frac{Z}{\Theta O}$, Maximum duration of granted
    upstream transm.\   \\
&   window size for Limited grant sizing \\
\hline
\end{tabular}
\end{table}

\begin{figure*}[t]
\begin{tabular}{c}
\setlength{\unitlength}{1mm}
\begin{picture}(160,42)
\thicklines
\put(20,10){\line(1,0){130}}
\put(20,30){\line(1,0){130}}
\put(20,5){\line(1,0){130}}
\put(20,31){\makebox(0,0)[b]{\scriptsize{OLT}}}
\put(20,9){\makebox(0,0)[t]{\scriptsize{ONU1}}}
\put(20,4){\makebox(0,0)[t]{\scriptsize{ONU2}}}

\put(29,30){\vector(1,-1){20}} 
\put(31,30){\vector(1,-1){25}} 
\put(49,10){\vector(1,1){20}} 
\put(57,10){\vector(1,1){20}} 
\put(56,5){\vector(1,1){25}} 
\put(70,5){\vector(1,1){25}} 
\put(83,30){\vector(1,-1){20}} 
\put(85,30){\vector(1,-1){25}} 
\put(103,10){\vector(1,1){20}} 
\put(114,10){\vector(1,1){20}} 

\thinlines
\thinlines
\put(50,11){\line(1,0){8}} 
\put(51,12){\line(1,0){8}}
\put(104,11){\line(1,0){11}} 
\put(105,12){\line(1,0){11}}
\put(57,6){\line(1,0){14}} 
\put(58,7){\line(1,0){14}}
\put(59,5){\line(0,1){2}}
\put(61,5){\line(0,1){2}}
\put(63,5){\line(0,1){2}}
\put(65,5){\line(0,1){2}}
\put(67,5){\line(0,1){2}}
\put(69,5){\line(0,1){2}}

\thinlines
\put(29,30){\line(0,1){3}}
\put(29,35){\makebox(0,0)[]{\scriptsize{$\gamma_{\alpha}(n-1,j=1,2)$}}}

\put(83,30){\line(0,1){5}}
\put(83,41){\makebox(0,0)[]{\scriptsize{$\gamma_{\alpha}(n,j = 1, 2) =$}}}
\put(83,38){\makebox(0,0)[]{\scriptsize{$\alpha(n-1,j=2) + t_R$}}}

\put(95,30){\line(0,1){3}}
\put(95,34.5){\makebox(0,0)[]{\scriptsize{$\Omega(n,j=1)$}}}

\put(123,30){\line(0,1){3}}
\put(123,34.5){\makebox(0,0)[]{\scriptsize{$\alpha(n,j=1)$}}}

\put(100,32){\vector(-1,0){5}}
\put(100,32){\vector(1,0){23}}
\put(109,33.25){\makebox(0,0)[]{\scriptsize{$I(n,j=1)$}}}

\put(42,21){\makebox(0,0)[]{\rotatebox{-45}{\scriptsize{grant messages}}}}

\put(63,21){\makebox(0,0)[]{\rotatebox{45}{\scriptsize{cycle $n-1$, $j= 1$}}}}
\put(74,18.5){\makebox(0,0)[]{\rotatebox{45}{\scriptsize{upstream transm.}}}}
\put(78,18.5){\makebox(0,0)[]{\rotatebox{45}{\scriptsize{cycle $n-1$, $j = 2$}}}}

\put(97.5,19){\makebox(0,0)[]{\rotatebox{-45}{\scriptsize{grant msgs.}}}}

\put(117,20){\makebox(0,0)[]{\rotatebox{45}{\scriptsize{upstream transm.}}}}
\put(121,20){\makebox(0,0)[]{\rotatebox{45}{\scriptsize{cycle $n$, $j = 1$}}}}
\end{picture} \\
\scriptsize{(a) Scheduling at the beginning: Report message included at the
   beginning (left side) of an upstream transmission}\\

\setlength{\unitlength}{1mm}
\begin{picture}(160,42)
\thicklines
\put(20,10){\line(1,0){130}}
\put(20,30){\line(1,0){130}}
\put(20,5){\line(1,0){130}}
\put(20,31){\makebox(0,0)[b]{\scriptsize{OLT}}}
\put(20,9){\makebox(0,0)[t]{\scriptsize{ONU1}}}
\put(20,4){\makebox(0,0)[t]{\scriptsize{ONU2}}}
\put(29,30){\vector(1,-1){20}} 
\put(31,30){\vector(1,-1){25}} 
\put(49,10){\vector(1,1){20}} 
\put(57,10){\vector(1,1){20}} 
\put(56,5){\vector(1,1){25}} 
\put(70,5){\vector(1,1){25}} 
\put(95,30){\vector(1,-1){20}} 
\put(97,30){\vector(1,-1){25}} 
\put(115,10){\vector(1,1){20}} 
\put(126,10){\vector(1,1){20}} 

\thinlines
\put(50,11){\line(1,0){8}} 
\put(51,12){\line(1,0){8}}
\put(116,11){\line(1,0){11}} 
\put(117,12){\line(1,0){11}}
\put(57,6){\line(1,0){14}} 
\put(58,7){\line(1,0){14}}
\put(59,5){\line(0,1){2}}
\put(61,5){\line(0,1){2}}
\put(63,5){\line(0,1){2}}
\put(65,5){\line(0,1){2}}
\put(67,5){\line(0,1){2}}
\put(69,5){\line(0,1){2}}

\thinlines
\put(29,30){\line(0,1){3}}
\put(29,35){\makebox(0,0)[]{\scriptsize{$\gamma_{\beta}(n-1,j=1,2)$}}}

\put(95,30){\line(0,1){3}}
\put(95,40){\makebox(0,0)[]{\scriptsize{$\gamma_{\beta}(n,j = 1, 2) =$}}}
\put(95,37){\makebox(0,0)[]{\scriptsize{$\beta(n-1,j = 2) = $}}}
\put(95,34){\makebox(0,0)[]{\scriptsize{$\Omega(n,j=1)$}}}

\put(135,30){\line(0,1){3}}
\put(135,34.5){\makebox(0,0)[]{\scriptsize{$\alpha(n,j=1)$}}}

\put(100,32){\vector(-1,0){5}}
\put(100,32){\vector(1,0){35}}
\put(115,33.25){\makebox(0,0)[]{\scriptsize{$I(n,j=1)$}}}

\put(42,21){\makebox(0,0)[]{\rotatebox{-45}{\scriptsize{grant messages}}}}

\put(63,21){\makebox(0,0)[]{\rotatebox{45}{\scriptsize{cycle $n-1$, $j= 1$}}}}
\put(74,18.5){\makebox(0,0)[]{\rotatebox{45}{\scriptsize{upstream transm.}}}}
\put(78,18.5){\makebox(0,0)[]{\rotatebox{45}{\scriptsize{cycle $n-1$, $j = 2$}}}}

\put(108,21){\makebox(0,0)[]{\rotatebox{-45}{\scriptsize{grant messages}}}}

\put(129,20){\makebox(0,0)[]{\rotatebox{45}{\scriptsize{upstream transm.}}}}
\put(133,20){\makebox(0,0)[]{\rotatebox{45}{\scriptsize{cycle $n$, $j = 1$}}}}

\end{picture} \\

\scriptsize{(b) Scheduling at the end:
Report message included at the
   end (right side) of an upstream transmission}
\end{tabular}
\caption{Illustration of cyclical report-grant polling structure.
Grant messages signal upstream transmission windows to the individual ONUs,
which report their queue depths in report messages included in the
upstream transmissions.
The figure also illustrates
scheduling at the beginning and at the end of an
upstream transmission for $O=2$ ONUs with offline
single-thread polling (STP).
For illustration of the differences in scheduling for cycle $n$,
scheduling decisions for cycle $n-1$ are assumed to be made
at the same time instant, namely at
$\gamma_{\alpha}(n-1, j = 1, 2) = \gamma_{\beta}(n-1, j = 1, 2)$.
With scheduling at the beginning, the report is at the beginning of
the upstream transmission, allowing the OLT to make the scheduling decision
for cycle $n$ at time instant
$\gamma_{\alpha}(n, j = 1, 2) = \alpha(n-1, j = 2) + t_R$, i.e., a report
transmission time $t_R$ after the upstream transmission
of ONU $j = 2$ of cycle $n-1$ begins to arrive at the OLT
(at $\alpha(n-1, j = 2)$).
With scheduling at the end, the report is at the end of the
upstream transmission, thus, the OLT can make the
scheduling decision for cycle $n$ at time instant
$\gamma_{\beta}(n, j = 1,2) = \beta(n-1, j = 2)$, i.e., when the
end of the upstream transmission of ONU $j = 2$ arrives at the OLT.}
\label{fig:cycle_illu}
\end{figure*}
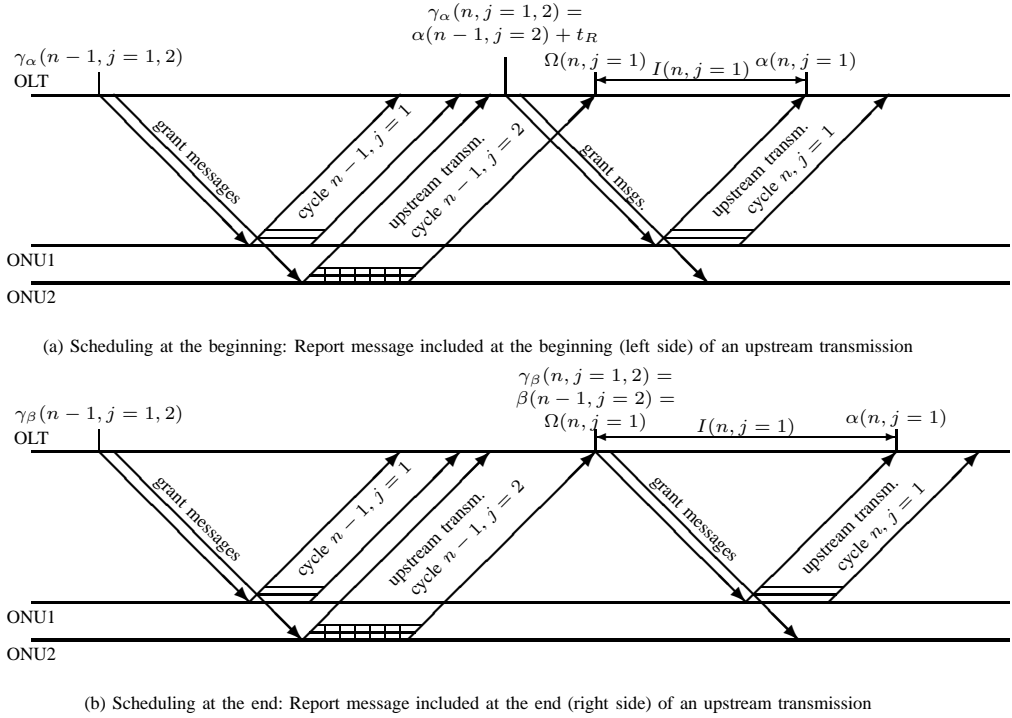
\subsection{Timing of Reporting at the Beginning and End of
Upstream Transmission} \label{time_beg_end:sec}
We initially consider two
report message scheduling approaches, namely reporting at the
beginning and reporting at the end of an upstream transmission. With
reporting at the beginning, the message indicating the queue depth
at the ONU to the OLT is positioned at the beginning of the upstream
transmission. Specifically, the report message that contains the
queue depth for sizing the upstream transmission of ONU $j$ of
thread $\theta$ in cycle $n$ begins to arrive at the OLT at time
instant $\alpha(n-1, \theta, j)$ and is completely received by time
instant $\alpha(n-1, \theta, j) + t_R$.
Thus, neglecting processing
delays, the OLT can make a scheduling decision based on this
received report as early as time instant $\alpha(n-1, \theta, j) +
t_R$, as illustrated for offline STP in Fig.~\ref{fig:cycle_illu}(a).
We denote $\gamma_{\alpha}(n, \theta, j)$ for the scheduling
instant for the upstream transmission of ONU $j$ of thread $\theta$
of cycle $n$ with reporting at the beginning, and specify
$\gamma_{\alpha}(n, \theta, j)$ for the different PON scheduling
frameworks~\cite{Mar01,ZhMo09} in Section~\ref{sch_beg_end:sec}.

In contrast, with reporting at the end, the report message
is positioned at the end of the upstream transmission,
i.e., it begins to arrive at the OLT at instant
$\beta(n-1, \theta, j) - t_R$ and is completely received by
instant $\beta(n-1, \theta, j)$.
Thus, the OLT can make grant sizing and scheduling decisions
for the upstream transmission of ONU $j$ of thread $\theta$ of cycle $n$
 as early as time instant $\beta(n-1, \theta, j)$,
as illustrated for offline STP in Fig.~\ref{fig:cycle_illu}(b).
We denote $\gamma_{\beta}(n, \theta, j)$ for the
scheduling instant for the upstream transmission of
ONU $j$ of thread $\theta$ of cycle $n$ with reporting at the end.

\subsection{Scheduling Instants with Reporting at Beginning and End}
\label{sch_beg_end:sec}
We consider the following combinations of
scheduling (polling) frameworks and grant sizing mechanisms:
\begin{itemize}
\item Offline single-thread polling with Gated grant sizing
    (S, offl., gat.)~\cite{kra01,KMP0202}
\item Offline single-thread polling with Limited grant
    distribution (S, offl., lim.)~\cite{kra01,KMP0202}
\item Offline single-thread polling with Excess grant distribution
         (S, offl., exc.)~\cite{AYDA1103,BAS0706}
\item Double Phase Polling with Excess grant distribution and share mechanism
    (D, exc shr.)~\cite{Sam01,Mar01}
\item Online single-thread polling with Limited grant distribution
          (S, onl. lim.)~\cite{kra01,KMP0202}
\item Online single-thread polling with Excess grant distribution
       (S, onl., exc.)~\cite{AYDA1103,BAS0706,MeMR13}
\item Online Multi-thread polling with Excess grant distribution
    (M, onl. exc.)~\cite{MeMR13}
\end{itemize}
With the offline scheduling (polling) framework, reports from
all $O$ ONUs must be received before the OLT makes
grant sizing and scheduling decisions.
Thus, the scheduling instant with S, offl. polling is
governed by the arrival of the report from the last ONU in a cycle,
i.e., for reporting at the beginning the scheduling instant
for the upstream transmission grants of a cycle $n$ coincides with
the arrival of the report message at the beginning of the upstream
transmission of the last ONU in the preceding cycle $n-1$,
$\gamma_{\alpha}(n, j) = \alpha(n-1, O) + t_R$,
as illustrated for $O = 2$ ONUs in Fig.~\ref{fig:cycle_illu}(a).
On the other hand, with reporting at the end, the OLT needs to wait until
the end of the upstream transmission of the last ONU in cycle $n-1$ is
received before sizing and scheduling the grants for cycle $n$, i.e.,
 $\gamma_{\beta}(n, j) = \beta(n-1, O)$, see Fig.~\ref{fig:cycle_illu}(b).

\begin{table}[t]
\caption{Scheduling instants $\gamma(n, \theta, j)$
for upstream transmissions of ONU $j$
(of thread $\theta$ in multi-thread polling) of cycle $n$}
\label{gam:tab}
\begin{tabular}{|l|c|c|c|} \hline
Scheduling  &    ONU                   & Rep. at the beg.  & Rep. at the end \\
 framework  & indices &         $\gamma_{\alpha}(n, \theta, j) = $
                                    & $\gamma_{\beta}(n, \theta, j) = $ \\
\hline
STP, offline  & $1 \leq j \leq O$ & $\alpha(n-1, O) + t_R$  & $\beta(n-1, O)$   \\
DPP   & $1 \leq j \leq \frac{O}{2}$  &   $\alpha(n-1, \frac{O}{2}) + t_R$  &
                                    $\beta(n-1, \frac{O}{2})$      \\
DPP   & $ \frac{O}{2} < j \leq O$  &   $\alpha(n-1, O) + t_R$  &
                                    $\beta(n-1, O)$      \\
STP, online  & $1 \leq j \leq O$  &   $\alpha(n-1, j) + t_R$  & $\beta(n-1, j) $   \\
MTP, online   & $1 \leq j \leq O$ & $\alpha(n-1, \theta, j) + t_R$   &  $\beta(n-1, \theta, j)$ \\
 \hline
\end{tabular}
\end{table}
Similarly, the scheduling instants of the other scheduling
frameworks depend on the arrival of the ONU report message
triggering the OLT grant sizing and scheduling either at the
beginning or end of the
ONU upstream transmission, as summarized in Table~\ref{gam:tab}.
Double-phase polling (DPP) schedules the first ONU group with
indices $j = 1, 2, \ldots, O/2$ when the report from ONU $O/2$ is
received and the second ONU group when the report from the last ONU
$O$ is received.
Online single-thread polling (STP) schedules each
individual ONU $j$ grant for a cycle $n$
immediately after receipt of the report from ONU $j$ in cycle $n-1$.
Similarly, online multi-thread polling schedules each ONU $j$ for a
given polling thread $\theta$ in a cycle $n$
immediately after receipt of the report
of ONU $j$ in thread $\theta$ of the preceding cycle $n-1$.

\subsection{Summary of Idle Time Analysis}
\label{Isum:sec}
In this section we summarize the analysis of the
channel idle time $I(n, \theta, j)$ that
precedes the arrival of the upstream transmission
of ONU $j$ of thread $\theta$ of cycle $n$ at the OLT,
which is detailed in Appendix~I.
The idle time $I(n, \theta, j)$ is the time span (period) from
the instant $\Omega(n, \theta, j)$ of the arrival of the end of the
preceding upstream transmission at the OLT
to the arrival of beginning of the upstream transmission of ONU $j$ of thread
$\theta$ in cycle $n$ at time instant $\alpha(n, \theta, j)$ at the
OLT, see Fig.~\ref{fig:cycle_illu}.
That is, the duration of the channel idle time is the difference
\begin{eqnarray}   \label{I:eqn}
 I(n, \theta, j) = \alpha(n, \theta, j) - \Omega(n, \theta, j).
\end{eqnarray}

The duration of this idle time span is governed by two constraints:
\begin{itemize}
\item Guard time constraint: There must be at least a guard time of
duration $t_g$ between
the arrival of two successive upstream transmissions at the OLT.
\item Signaling constraint: The upstream transmission of ONU $j$
of thread $\theta$
of cycle $n$ can arrive no earlier than the gate signaling delay
$T(n, \theta, j)$ (transmission time of grant message $t_G$ plus
round-trip propagation delay $2 \tau$) after the scheduling instant
$\gamma(n, \theta, j)$.
\end{itemize}
As detailed in Appendix~I, the earlier scheduling instant
$\gamma_{\alpha}(n, \theta, j)$
with reporting at the beginning
compared to $\gamma_{\beta}(n, \theta, j)$ for reporting at the end
can reduce the channel idle time.

Depending on the combination of guard time and signaling constraints
that govern the idle time for the reporting at the beginning and end,
reporting at the beginning can reduce the idle time up to the
difference between the two scheduling instants, i.e.,
up to $\gamma_{\beta}(n, \theta, j) - \gamma_{\alpha}(n, \theta, j)$.

\subsection{Dynamic Optimization of Report Message Scheduling}
\label{opt:sec}
The report message scheduling (RMS) can be
dynamically selected for optimization.
Reporting at the end (and thus including the
packets that have been newly generated during an upstream transmission
in the report) can be dynamically selected when reporting at the
beginning would not reduce the channel idle time.
Based on the detailed analysis in Appendix~I,
the idle time with offline polling hinges primarily on the reporting of the
last ONU ($j = O$) in a cycle.
Thus, all but the last ONU, i.e., ONUs $j = 1, 2, \ldots, O-1$,
can report at the end, thus including the newly generated packets in the
report, while the last ONU $j = O$ reports at the beginning.

For online scheduling, RMS dynamic selection is not possible.
This is because the report schedule decision (beginning or end reporting)
would need to be communicated by the OLT to the ONU before the
parameters determining the channel idle time reduction $\Delta_I$
case in Table~\ref{DI:tab} are available at the OLT.
In particular, to impact the idle time preceding the ONU $j$ transmission
arrival in cycle $n$, the ONU would need to be instructed to report
at the beginning or end of the upstream transmission in cycle $n-1$
(when the report determining ONU $j$'s upstream transmission window
in cycle $n$ arrives at the OLT).
The OLT would need to send out these instructions for reporting
at the beginning or end in the preceding cycle $n-2$.
However, the queue depth of the preceding ONU $j-1$ used for sizing
ONU $j-1$'s window in cycle $n$ (which governs $\Omega(n, j)$)
is not yet available at the OLT at that time (as it arrives only
shortly before the report from ONU $j$ in cycle $n-1$).
Thus, the report scheduling cannot be optimized unless some
traffic prediction~\cite{LuAn05,ZM0708} is employed.

\section{Simulation Results for Packet Delay}
\label{perf:sec}

\subsection{Simulation Set-up}
We employ a simulation model of the OLT and ONUs built on CSIM, a discrete
event simulator using the C programming language,
and validated in preceding studies~\cite{Mar01,MeMR13}.
We implement the LRPON in both EPON
and GPON standards for $C = 1$~Gbps (IEEE 802.3ah and G.984,
respectively) and $C=10$~Gbps (IEEE 802.3av and G.987, respectively),
with a total number of $O=32$ infinite-buffer
ONUs (ONTs in GPON)  placed around the OLT
with a constant distance of $90$~km from the OLT to the splitter and the
ONUs placed randomly in the last $10$~km range. The
maximum round-trip delay is $2 \tau = 1$~ms.

We consider self-similar packet traffic with Hurst parameter 0.75
and four different packet sizes with distribution 60\% 64 Byte, 4\%
300 Byte, 11\% 580 Byte, and 25\% 1518 Byte packets. The traffic
load is defined as long-run average of the payload bit rate.

Control messages for EPON and GPON follow the respective standards.
In GPON, the control message is sent periodically every
125~$\mu$sec.
In the EPON, the ONUs report their queue depths with a REPORT message
(64~Bytes), while a DBRu (4~Bytes) message is used in the GPON.
We set the guard times for EPON $t_g = 1~\mu$s and for GPON $t_g = 30$~ns.

Simulations are performed for all DBAs noted in
Section~\ref{sch_beg_end:sec} for
maximum cycle length $Z = 2$, 4, and 8~ms.
The maximum grant size for limited grant sizing~\cite{kra01,KMP0202},
which is the initial basis for excess bandwidth
allocation~\cite{AYDA1103,  BAS0706} is
\begin{eqnarray} \label{Gmax:eqn}
G_{\max} = \frac{Z}{\Theta O}.
\end{eqnarray}
For MTP, we set the number of threads to $\Theta=2$ (for consistent
comparison with the two ONU groups in DPP~\cite{Sam01}) and the
threshold for thread tuning to $T_{\rm tune} = 5$~\cite{Muk01,MeMR13}.

We observe the average
packet delay from the packet generation instant at an ONU
to the delivery instant of the complete packet to the OLT.
We also observe the average channel idle time $I(n, \theta, j)$.

\begin{figure}[t!]
\centering
\begin{tabular}{c}
\includegraphics[scale=0.65]{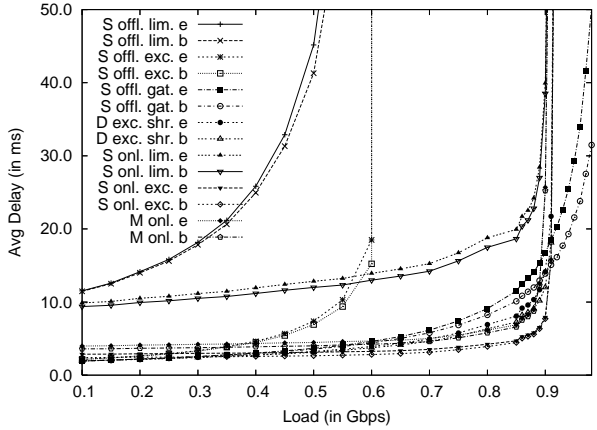} \\
\footnotesize{a) Max. cycle length $Z=2$~ms} \\
\includegraphics[scale=0.65]{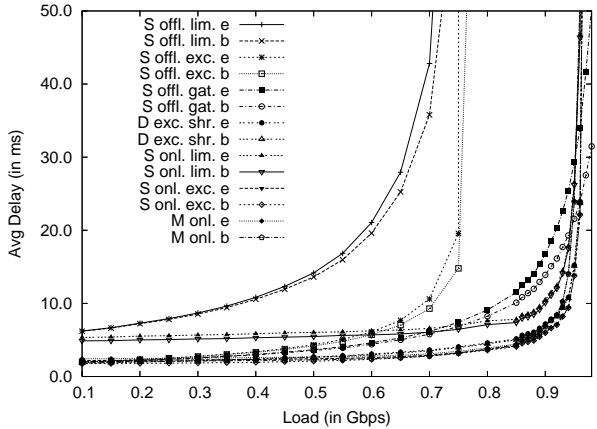} \\
\footnotesize{b)  Max. cycle length $Z=4$~ms} \\
\includegraphics[scale=0.65]{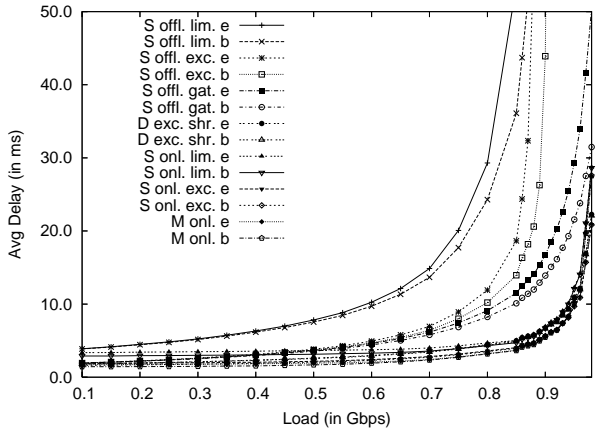} \\
\footnotesize{c)  Max. cycle length $Z=8$~ms}\\
\end{tabular}
\caption{Mean packet delay for EPON with upstream bandwidth $C = 1$~Gbps.
Abbreviations for DBA mechanisms (see Section~\ref{sch_beg_end:sec}):
Threads: S = single-thread polling, D = double-phase polling,
M = multi-thread polling;
Scheduling framework: offl. = offline, onl. = online;
Grant sizing: lim. = limited, exc. = excess distribution, gat. = gated,
exc. shr = excess share;
Report scheduling: e = end, b = beginning.}
\label{fig:OD-1G-EPON}
\end{figure}
\begin{figure}[t!]
\centering
\includegraphics[scale=0.65]{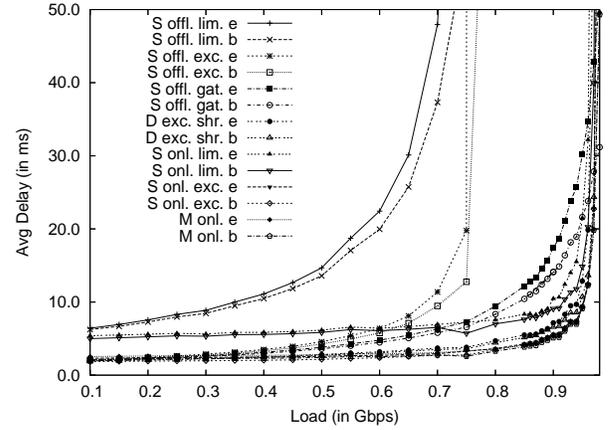}
\caption{Mean packet delay for xGPON with $C=1$~Gbps and maximum
cycle length $Z = 4$~ms.}
\label{fig:OD-1G-GPON}
\end{figure}
In Figs.~\ref{fig:OD-1G-EPON} and~\ref{fig:OD-10G-EPON} we plot the
average packet delay for all considered combinations of scheduling
framework and grant sizing mechanism (see
Section~\ref{sch_beg_end:sec}) for all three considered maximum
cycle lengths $Z$ for the 1G and 10G EPON respectively. The
corresponding average channel idle times are plotted in
Figs.~\ref{fig:IT-1G-EPON} and~\ref{fig:IT-10G-EPON}. A few
scheduling framework-grant sizing combinations were omitted from
Figs.~\ref{fig:IT-1G-EPON} and~\ref{fig:IT-10G-EPON} to reduce
clutter. Specifically, for reporting at the end, all offline STP
approaches give essentially the same average idle times; we plot
therefore only offline STP with gated grant sizing while omitting
offline STP with limited and excess grant sizing.
Moreover, for reporting at the beginning, offline STP
with limited grant sizing gives very similar average idle times to
offline STP with excess grant distribution (S. offl. exc.);
therefore, we only plot S. offl. exc. We omitted online STP with
excess grant distribution (S. onl. exc.) which gives very similar
average idle times as online STP with limited grant sizing (S. onl.
lim.). Due to space constraints, we include for the 1G and 10G GPON
only the simulation results for the representative $Z = 4$~ms
maximum cycle length in
Figs.~\ref{fig:OD-1G-GPON},~\ref{fig:IT-1G-GPON},~\ref{fig:OD-10G-GPON},
and~\ref{fig:IT-10G-GPON}.

\begin{table}
\caption{Impact of number of ONUs $O$: Average packet delay,
idle time per ONU transmission $I$, cycle length,
and ONU transmission window
length $\bar{G}$ for $O = 8$ and 32 ONUs;
fixed parameters: $C = 1$~Gbps EPON,
STP offline gated DBA, traffic load = 0.9~Gbps.}
\label{O:tab}
\begin{tabular}{|l|ccc|ccc|} \hline
  & \multicolumn{3}{|c|}{$O = 8$}&\multicolumn{3}{|c|}{$O = 32$}\\
Perf. Metric & end & beg.  & opt.          & end  & beg. &  opt. \\ \hline
Avg. pkt. del. [ms] & 15.1  & 8.8 & 8.7  & 16.7  & 13.9  & 13.0 \\
Avg. idl. tim. [$\mu$s] & 123 & 67.5 & 61.9 &  31.6 & 25.9  & 22.6\\
Avg. cyc. len. [ms]   & 8.2 & 4.5 & 5.0 & 8.0  & 7.1 &  7.7 \\
Avg win. len. $\bar{G}$ [ms]  & 0.90  & 0.49  & 0.55  &  0.22  & 0.19  & 0.20 \\
\hline
\end{tabular}
\end{table}

\subsection{General Reporting at Beginning vs. End Trade-off}
We observe across the set of plots in
Figs.~\ref{fig:OD-1G-EPON}--\ref{fig:IT-10G-GPON}
that reporting at the beginning generally gives lower
average packet delays and channel idle times than reporting
at the end.
That is, the effect of the OLT receiving reports earlier
with reporting at the beginning and thus making earlier
upstream transmission sizing and scheduling decisions generally
outweighs the effect of reporting the newly generated packets
(generated during an ONU upstream transmission) later (i.e., in the next cycle).
The earlier reporting tends to reduce the channel idle time and thus
increases the level of masking of idle times, resulting in overall
shorter polling cycles and thus lower packet delays.
The specific delay reduction effects for the various DBA
mechanisms are discussed in detail in the following subsections.

Before examining the individual DBA mechanisms, we illustrate
the effect of the number of ONUs $O$ on the impact of
report scheduling.
In Table~\ref{O:tab}, we consider STP with offline gated DBA in
a $C = 1$~Gbps EPON at traffic load of 0.9~Gbps.
We observe from the table that for a low number of $O = 8$ ONUs,
reporting at the beginning reduces the average packet
delay almost to half the delay
for reporting at the end; whereas for the higher number of $O = 32$ ONUs,
the delay reduction with reporting at the beginning is far less pronounced.
For the smaller number of ONUs, each ONU upstream
transmission window constitutes a relatively larger portion of the
overall cycle duration, as illustrated by
the average cycle length
and average ONU transmission window length values $\bar{G}$
in Table~\ref{O:tab}.

For reporting at the end, the offline DBA considered in
Table~\ref{O:tab} has a $2 \tau$
channel idle period between successive cycles~\cite{ZhMo09}.
Thus, neglecting the guard times $t_g$ and the small variations in the
round-trip propagation delays, the average idle time is
approximately $2 \tau/O$.

Reporting at the beginning masks a portion of this propagation delay
equal to the length of the last transmission window in a cycle.
Thus, the average idle time is reduced to roughly
$(2 \tau - \bar{G})/O$.
With each transmission window (including the last window in the cycle)
constituting a relatively larger portion of the cycle for small $O$,
this masking effect due to reporting at the beginning is significantly
more pronounced for small $O$ than for large $O$.
The relatively stronger masking effect for small $O$
leads to significantly more pronounced shortening of the
average cycle duration and the average channel idle time, and, in turn,
the average packet delay.
To summarize, the performance improvements with reporting
at the beginning generally are more pronounced in PONs with
small numbers of ONUs.
However, for the current trend of increasing
numbers of ONUs served in a PON, the impact of report scheduling
is reduced.

We observe from the results  in the
``opt.'' columns in Table~\ref{O:tab} that
the quantitative benefits from dynamic optimization
of the report message scheduling (see Section~\ref{opt:sec})
are generally small.
Including the newly generated packets in the reporting at the
end slightly increases the average transmission window length and
cycle length as more packets are included in the ONU reports
sent at the end of an upstream transmission.
The overall longer cycle length increases also the window
of the last ONU, thus increasing it in proportion to the
round-trip propagation delay and, in turn, reducing average idle time
compared to reporting at the beginning.
The combined effects of including the newly generated packets in the
end reports and the reduced idle time reduce the average packet delay.
It is important to keep in mind though that these effects are
relatively small, and this optimization through dynamic RMS selection
is limited to offline scheduling.

\subsection{Offline Single-thread Polling (STP) with Limited
Grant Sizing (S offl. lim.)}
We observe from the packet delay plots in
Figs.~\ref{fig:OD-1G-EPON},~\ref{fig:OD-1G-GPON},~\ref{fig:OD-10G-EPON},
and~\ref{fig:OD-10G-GPON} that offline STP with limited grant sizing
gives the highest average packet delay among the compared
DBA approaches. This is mainly due to the strict limit
$G_{\max}$ on the transmission window length per ONU in a cycle, which
results in inflexible bandwidth allocation to the individual ONUs.
Offline STP with reporting at the end
utilizes a maximum portion of $Z/(2 \tau + Z)$ of a cycle
for upstream transmissions since the upstream channel is idle during the
upstream propagation of the last report of a cycle and downstream
propagation of the first grant of the next cycle.
That is, from the OLT perspective,
the gate signalling delay $T$ from the scheduling instant to
the arrival of the corresponding upstream transmission at the OLT
is roughly the round-trip propagation delay $2 \tau$ (when
neglecting the small gate message transmission times).

Examining
Figs.~\ref{fig:OD-1G-EPON},~\ref{fig:OD-1G-GPON},~\ref{fig:OD-10G-EPON},
and~\ref{fig:OD-10G-GPON} closer for the impact of reporting at the
end vs. reporting at the beginning, we observe relatively small
delay differences for the short $Z=2$~ms maximum cycle length. For
the longer $Z = 4$~ms and 8~ms cycle lengths, we observe substantial
delay reductions with reporting at the beginning at high traffic
loads.
These delay reductions can be explained with the average
channel idle times plotted in
Figs.~\ref{fig:IT-1G-EPON},~\ref{fig:IT-1G-GPON},~\ref{fig:IT-10G-EPON},
and~\ref{fig:IT-10G-GPON}, as discussed jointly with
offline STP with excess bandwidth allocation in
the next section.

\subsection{Offline STP with Excess Bandwidth Allocation (S offl. exc.)}
\label{Sofflexc:sec}
Excess bandwidth allocation~\cite{AYDA1103,BAS0706} makes the dynamic bandwidth
allocation to the individual ONUs more flexible by redistributing
the unused portions of the $G_{\max}$ limit from ONUs with presently
low traffic to ONUs that presently have large traffic queues.
As a result, the polling cycles become better utilized, which results in
substantial delay reductions compared to limited grant sizing,
as observed in
Figs.~\ref{fig:OD-1G-EPON},~\ref{fig:OD-1G-GPON},~\ref{fig:OD-10G-EPON},
and~\ref{fig:OD-10G-GPON}.

\begin{figure}[t!]
\centering
\begin{tabular}{c}
\includegraphics[scale=0.65]{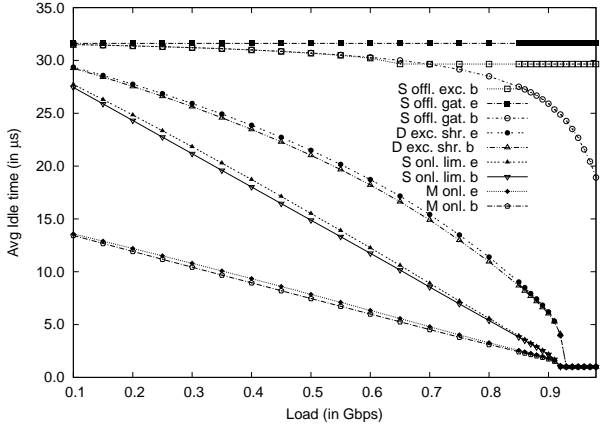} \\
\footnotesize{a) Max. cycle length $Z=2$~ms} \\
\includegraphics[scale=0.65]{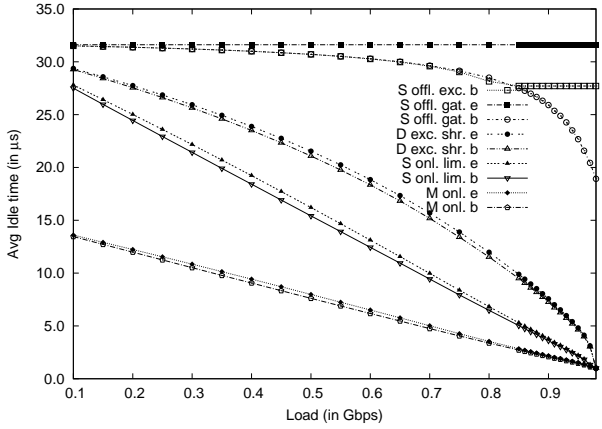} \\
\footnotesize{b) Max. cycle length $Z=4$~ms} \\
\includegraphics[scale=0.65]{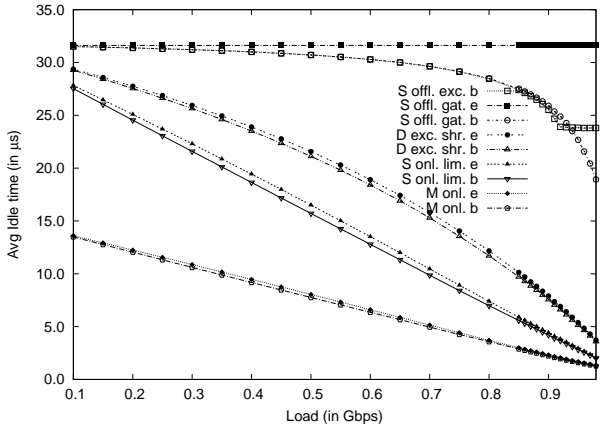} \\
\footnotesize{c) Max. cycle length $Z=8$~ms} \\
\end{tabular}
\caption{Mean duration of channel idle time per ONU
upstream transmission for EPON with upstream bandwidth $C=1$~Gbps.}
\label{fig:IT-1G-EPON}
\end{figure}

\begin{figure}[t!]
\includegraphics[scale=0.65]{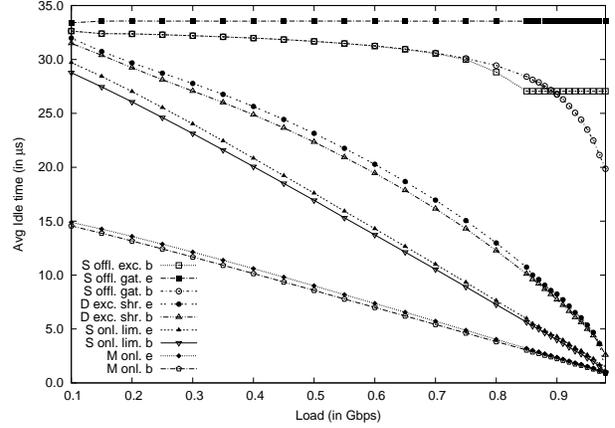}
\caption{Mean duration of channel idle time per ONU
upstream transmission for xGPON for $C=1$~Gbps and maximum
cycle length $Z = 4$~ms.}
\label{fig:IT-1G-GPON}
\end{figure}

As the traffic load increases, we observe
from Figs.~\ref{fig:IT-1G-EPON},~\ref{fig:IT-1G-GPON},~\ref{fig:IT-10G-EPON},
and~\ref{fig:IT-10G-GPON}
reductions in the average idle time for offline
STP with excess grant sizing and reporting at the beginning
compared to offline STP with gated grant sizing with reporting at the end
(which is plotted as a representative for all offline STP approaches with
reporting at the end).
As noted above and elaborated in more detail in Appendix~I,
with offline STP, there is a mandatory
$2 \tau$ idle time between successive cycles.
Thus, the arrival of the first ONU ($j=1$) transmission
at the OLT is preceded
by a $2 \tau$ idle time, while the subsequent ONU
transmissions ($j = 2, 3, \ldots, 32$) within the cycle
are preceded by a guard time $t_g$ (provided the traffic load
and resulting grant lengths are sufficient to mask the
propagation delay differences~\cite{Mar03}).
With reporting at the end, the average idle time per ONU
transmission is thus approximately $2 \tau/O \approx 31.25~\mu$s,
where we neglect the $t_g$ guard times and consider $2 \tau = 1$~ms.
With reporting at the beginning of the upstream transmission, the
idle time is reduced by the length of the upstream transmission of the
last ONU in a cycle, which approaches $G_{\max}$ with high traffic load.
Thus, the average idle time is reduced to $(2\tau - G_{\max})/O$,
which is approximately $27.3~\mu$s for $Z = 4$~ms.
This reduced average channel idle time per ONU upstream transmission
reduces the average packet delay and increases the utilization
of the upstream channel.

In additional idle time evaluations which we do not include in the
plots to avoid clutter, we found that the differences between
reporting at the beginning and reporting
at the end of an upstream transmission are very similar for
limited grant sizing and for excess grant sizing.
The main difference between limited and excess grant sizing
is that the average ONU transmission is longer with excess
grant sizing, which results in the lower delays
observed in
Figs.~\ref{fig:OD-1G-EPON},~\ref{fig:OD-1G-GPON},~\ref{fig:OD-10G-EPON},
and~\ref{fig:OD-10G-GPON}.
However, for very high traffic loads with delays beyond the plotted range,
both limited and excess grant sizing exhibit the same respective
utilization limits of $Z/(2 \tau + Z)$ with reporting
at the end and $(Z - G_{\max})/(2\tau + Z)$ with reporting in the
beginning.

\subsection{Offline STP with Gated Grant Sizing (S offl, gat.)}
Gated grant sizing does not limit the lengths of the ONU upstream transmission
windows.
Thus, for high traffic loads, the window lengths grow very large,
substantially larger than $G_{\max}$.
Accordingly, we observe in
Figs.~\ref{fig:IT-1G-EPON},~\ref{fig:IT-1G-GPON},~\ref{fig:IT-10G-EPON},
and~\ref{fig:IT-10G-GPON}
a substantially more pronounced reduction of the average channel idle time
per ONU transmission for reporting at the beginning with gated
grant sizing than with excess grant sizing.

\begin{figure}[t!]
\centering
\begin{tabular}{c}
\includegraphics[scale=0.65]{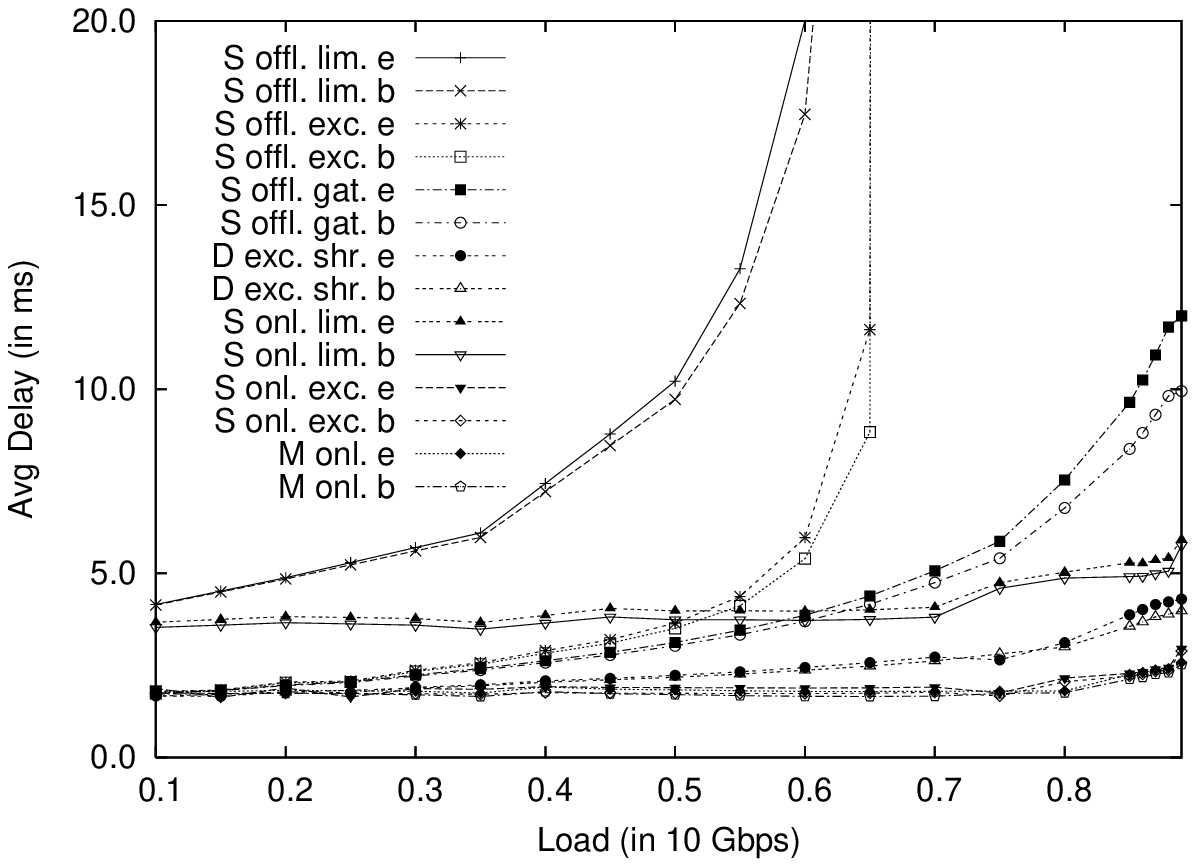} \\
\footnotesize{a) Max. cycle length $Z=2$~ms} \\
\includegraphics[scale=0.65]{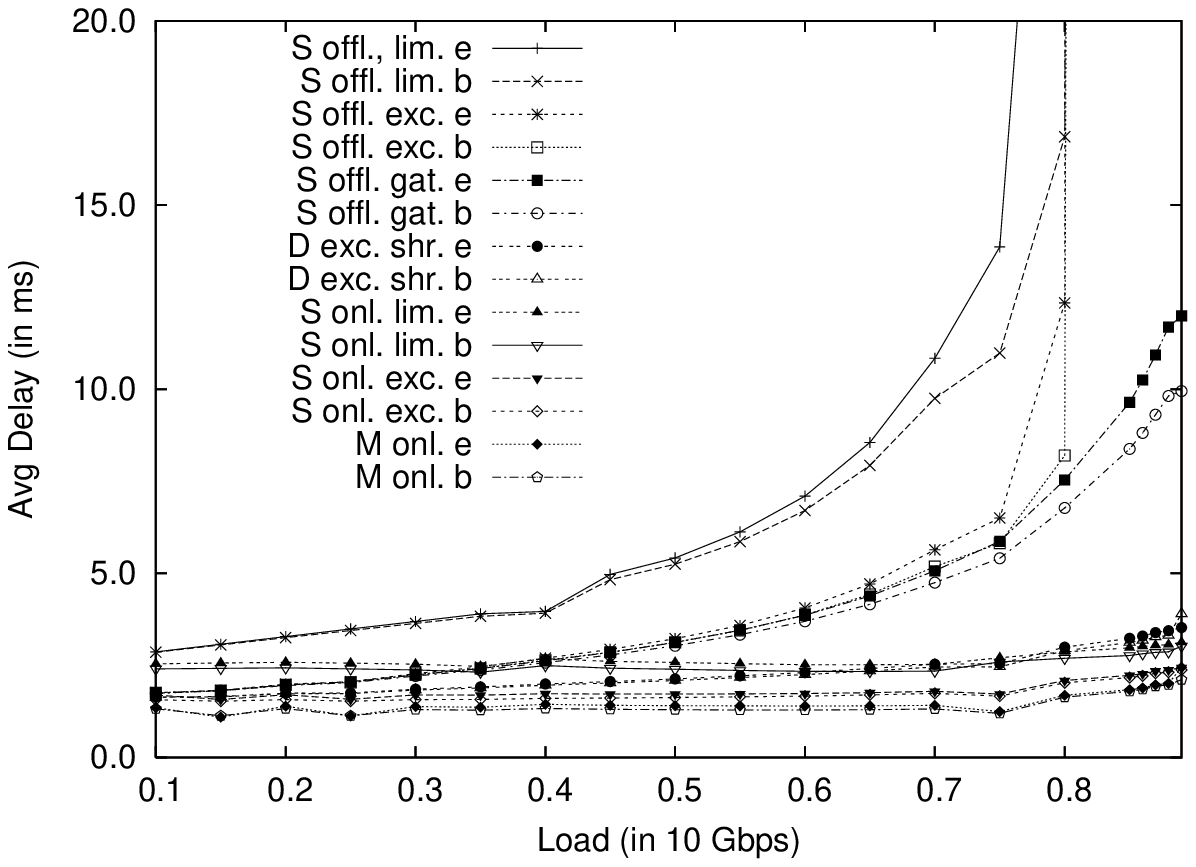} \\
\footnotesize{b) Max. cycle length $Z=4$~ms} \\
\includegraphics[scale=0.65]{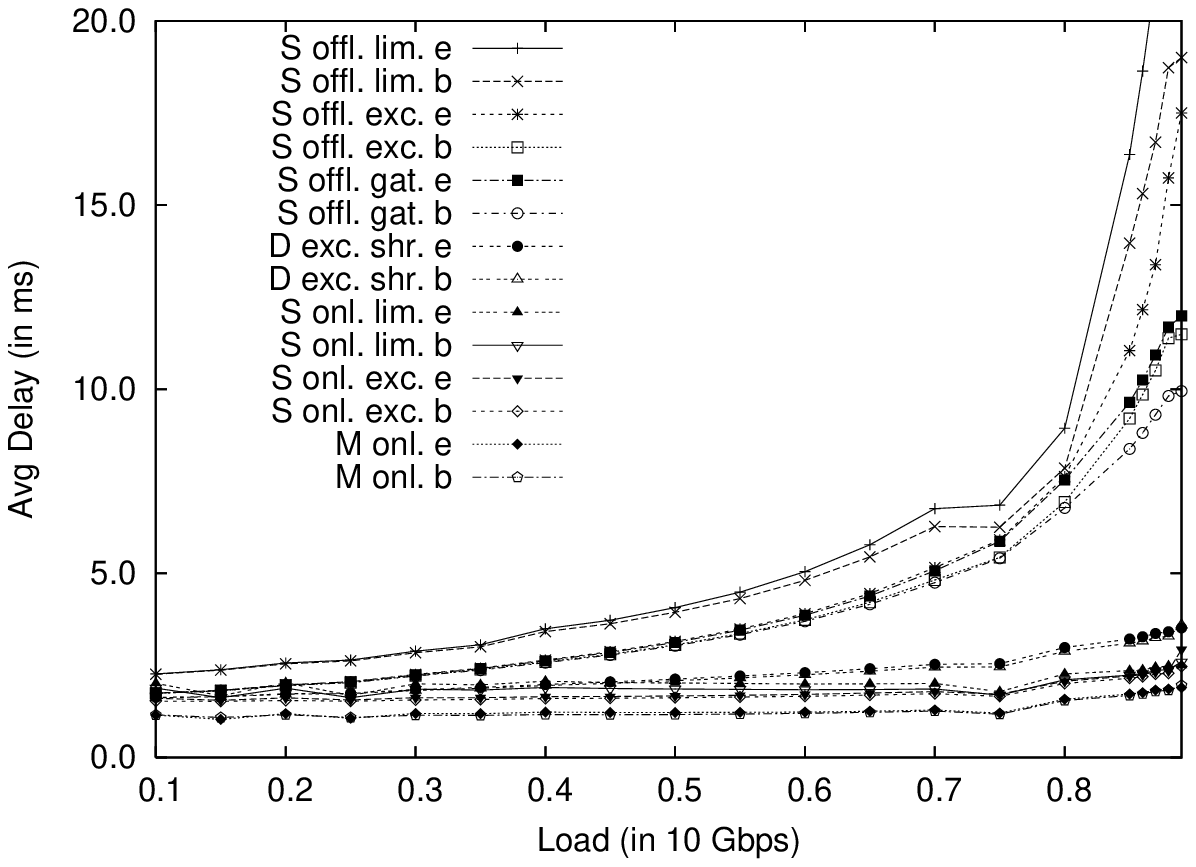} \\
\footnotesize{c) Max. cycle length $Z=8$~ms}
\end{tabular}
\caption{Mean packet delay for EPON with upstream bandwidth $C=10$~Gbps.}
\label{fig:OD-10G-EPON}
\end{figure}

\begin{figure}[t!]
\centering
\includegraphics[scale=0.65]{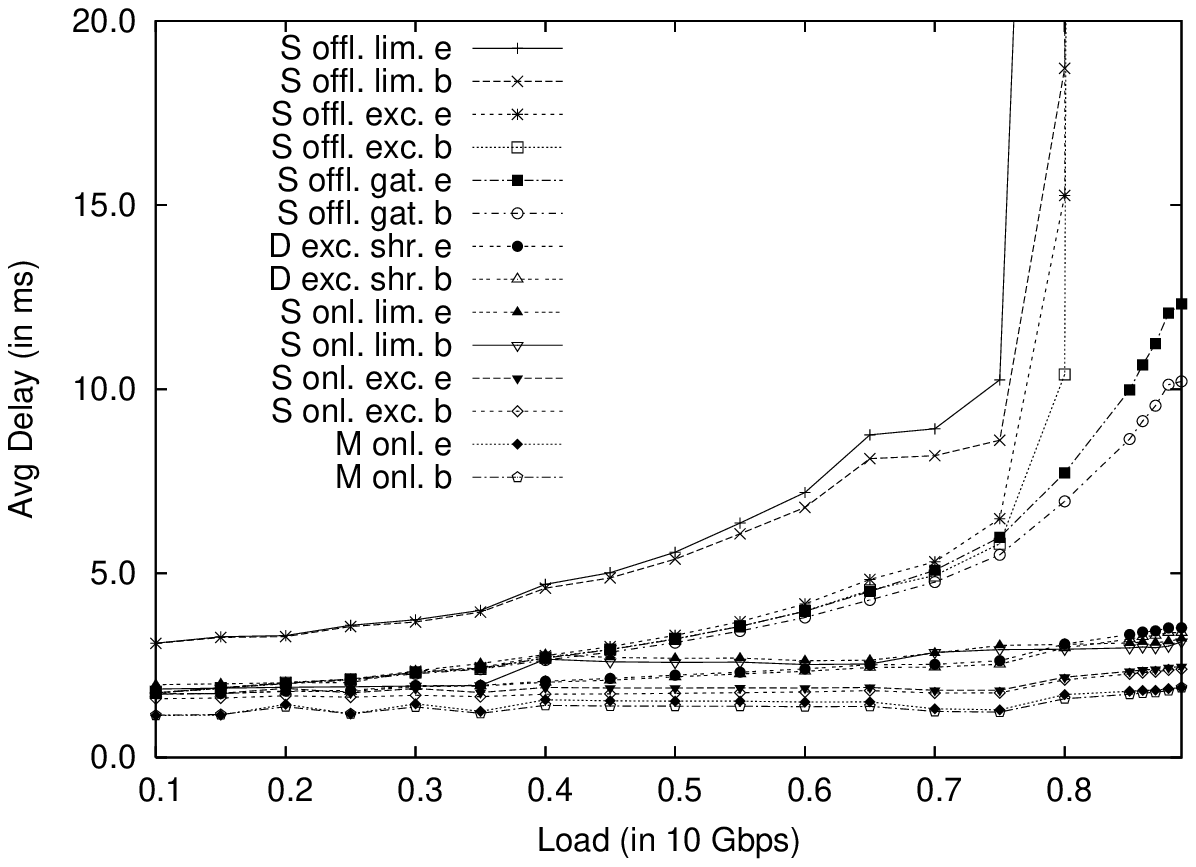}
\caption{Mean packet delay for xGPON with $C=10$~Gbps and
maximum cycle length $Z = 4$~ms.}
\label{fig:OD-10G-GPON}
\end{figure}

Correspondingly, we observe in
Figs.~\ref{fig:OD-1G-EPON},~\ref{fig:OD-1G-GPON},~\ref{fig:OD-10G-EPON},
and~\ref{fig:OD-10G-GPON} relatively large reductions of the average
packet delay with reporting at the beginning compared to reporting
at the end. The delay reduction reaches about 20~ms at the 0.98~Gbps
load point in Figs.~\ref{fig:OD-1G-EPON} and~\ref{fig:OD-1G-GPON}.

\subsection{Online STP with Limited and Excess Grant Sizing
(S onl. lim. and S onl. exc.)}
\begin{figure}[t!]
\centering
\begin{tabular}{c}
\includegraphics[scale=0.65]{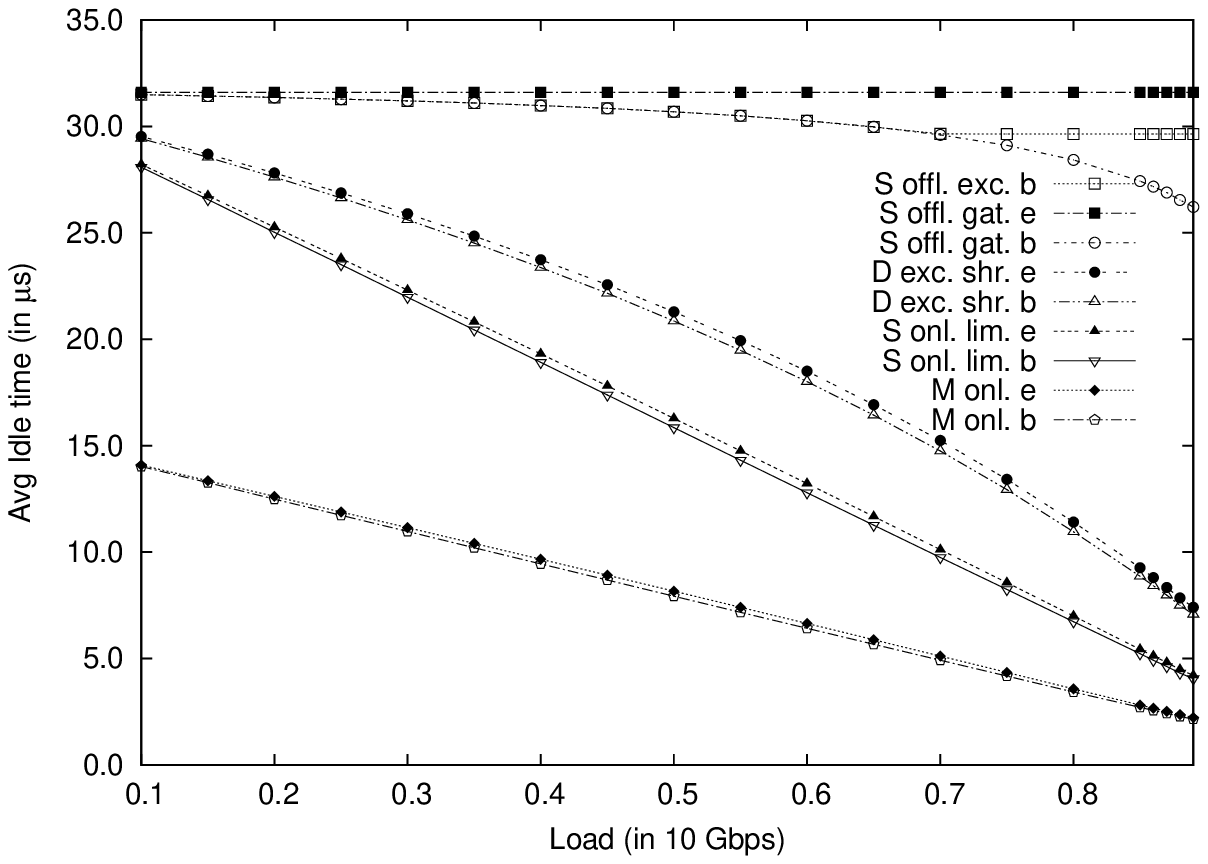} \\
\footnotesize{a) Max. cycle length $Z=2$~ms} \\
\includegraphics[scale=0.65]{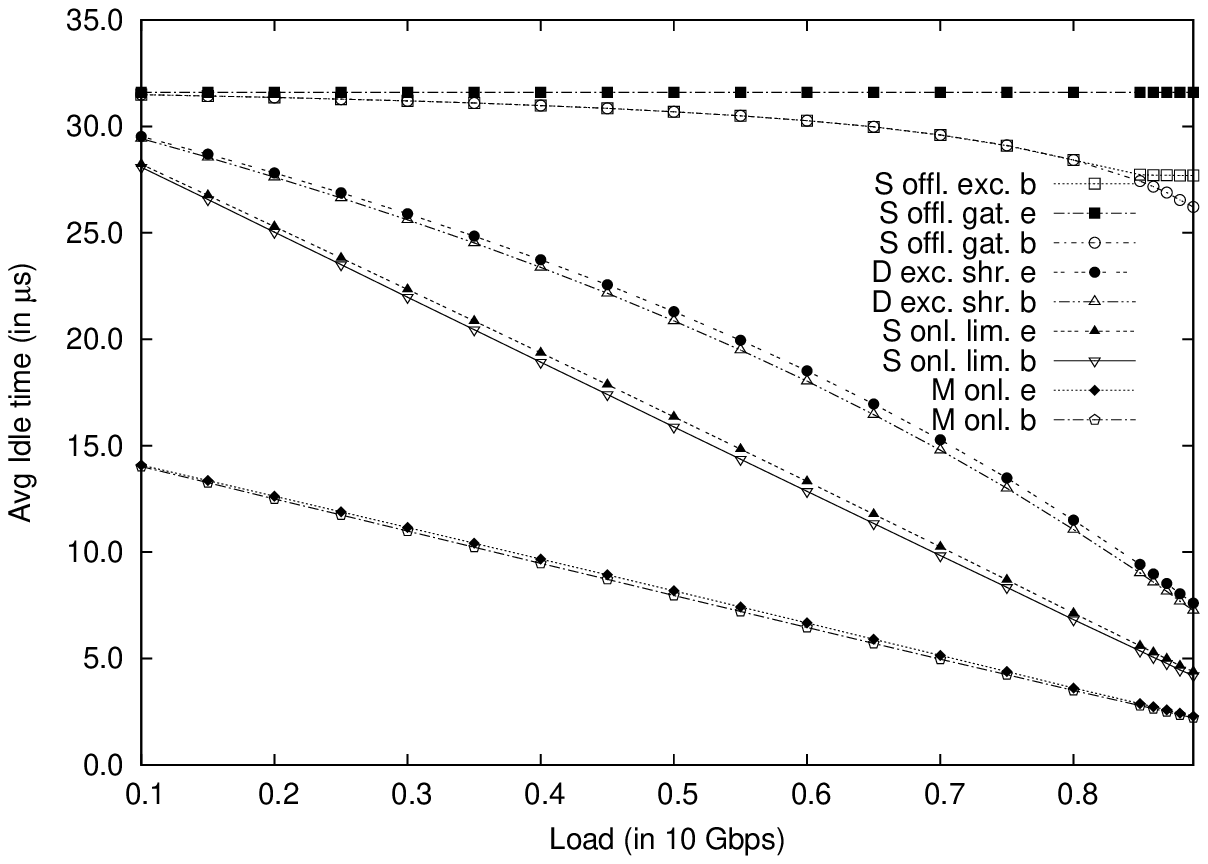} \\
\footnotesize{b) Max. cycle length $Z=4$~ms} \\
\includegraphics[scale=0.65]{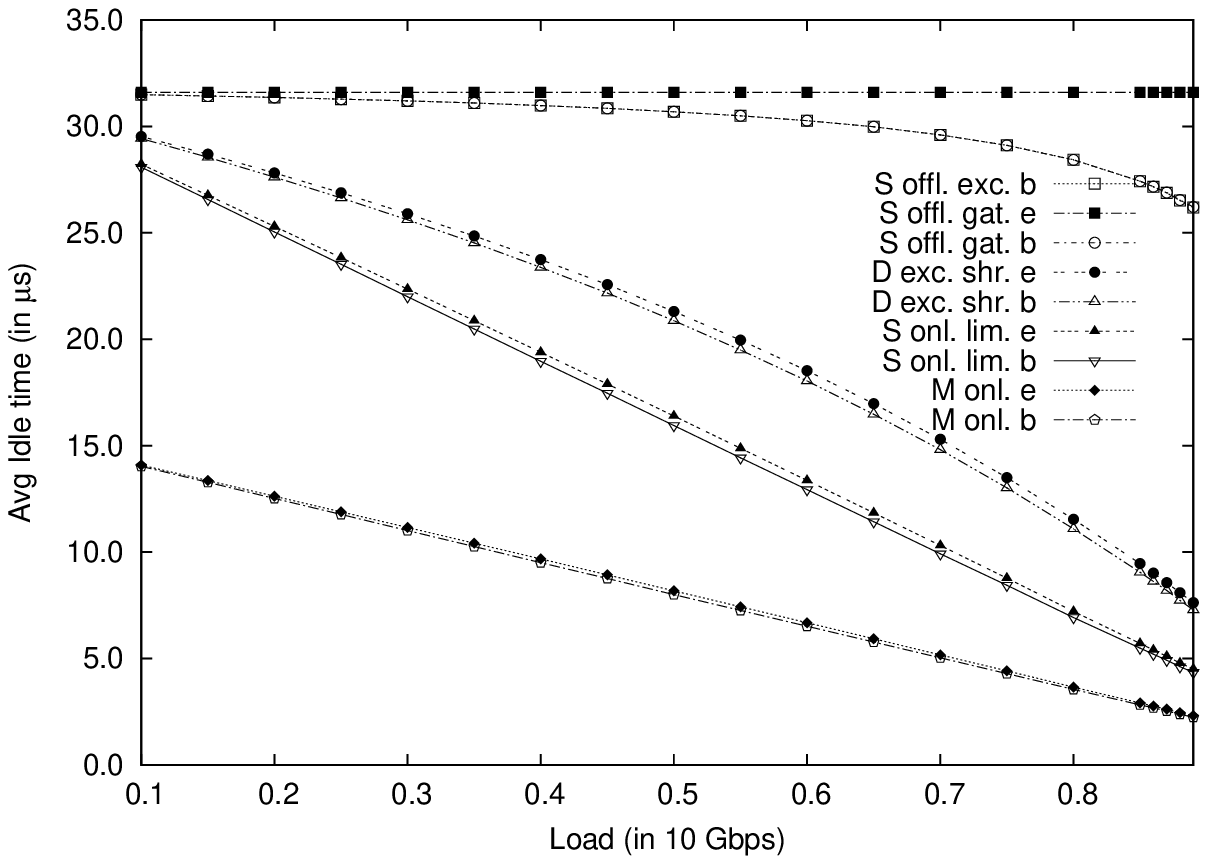} \\
\footnotesize{c) Max. cycle length $Z=8$~ms} \\
\end{tabular}
\caption{Mean duration of channel idle time per ONU upstream transmission
 for EPON with bandwidth $C=10$~Gbps.}
\label{fig:IT-10G-EPON}
\end{figure}

\begin{figure}[t!]
\centering
\includegraphics[scale=0.65]{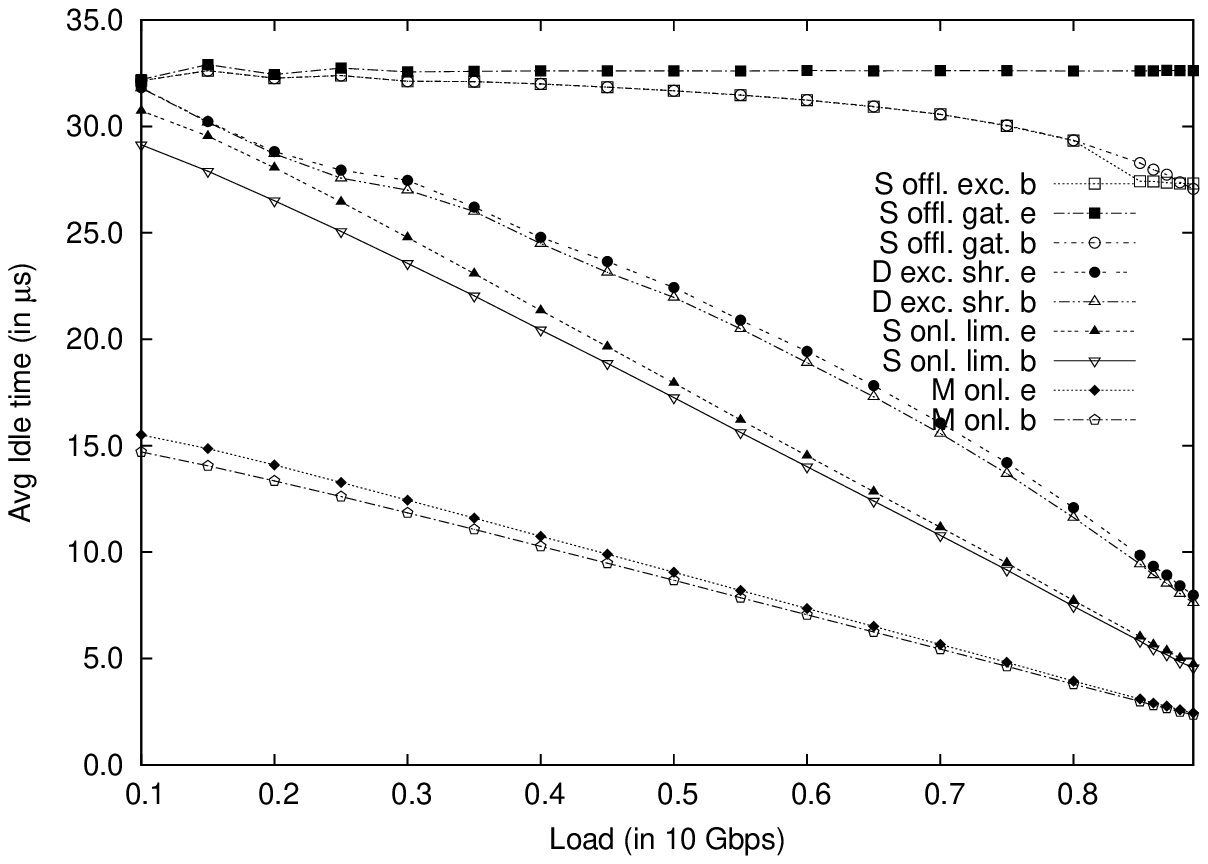}
\caption{Mean duration of channel idle time per ONU upstream transmission
   for xGPON for $C=10Gbps$ and maximum cycle duration $Z = 4$~ms.}
\label{fig:IT-10G-GPON}
\end{figure}

We observe from Figs.~\ref{fig:OD-1G-EPON}--\ref{fig:IT-10G-GPON}
that
(a) online STP with excess grant sizing gives substantially
smaller delays than online STP with limited grant sizing,
and (b) reporting at the beginning gives only very minuscule
reductions (on the order of 1--3~ms)
in delay compared to reporting at the
end for these two DBA approaches.
The advantage of excess grant sizing is again due to the more
flexible transmission window allocations to the individual ONUs,
which more quickly serves their bursty traffic.

By closely examining the online STP with limited grant sizing
delay performance across
Figs.~\ref{fig:OD-1G-EPON},~\ref{fig:OD-1G-GPON},~\ref{fig:OD-10G-EPON},
and~\ref{fig:OD-10G-GPON}, we observe delay reductions with
increasing maximum cycle length $Z$ and upstream bandwidth $C$.
For instance, we observe from Fig.~\ref{fig:OD-1G-EPON} for
$C = 1$~Gbps that the average packet delay at traffic load 0.8
is close to 18~ms for $Z = 2$~ms, but drops to around 7.5~ms
for $Z = 4$~ms and further to roughly 4.5~ms for $Z = 8$~ms.
Similarly, comparing Fig.~\ref{fig:OD-1G-EPON}a) with
Fig.~\ref{fig:OD-10G-EPON}a) for $Z= 2$~ms,
 we observe that the higher $C = 10$~Gbps bandwidth reduces the
average packet delay to less than half of the delays for $C =
1$~Gbps. These observed delay reductions are due to the increased
limit on the ONU upstream transmission $G_{\max}$ (\ref{Gmax:eqn}),
which increases the flexibility of the dynamic bandwidth allocation
of limited grant sizing. In particular, we observe from
Fig.~\ref{fig:OD-10G-GPON}c) that for the largest considered
$G_{\max}$, limited grant sizing attains essentially the same
average delays as excess grant sizing.
Also, the higher channel bit rate reduces the
relative impact (in terms of time delay) due to the fixed-size
(in terms of Byte count) overheads.

Online STP interleaves the polling processes to the individual ONUs
(with a single polling process per ONU), eliminating the
$2 \tau$ idle period between successive cycles in offline polling.
Consequently, there are fewer and smaller opportunities for
reducing unmasked idle time by shifting the report message from
the end to the beginning of the upstream transmission,
as validated by the idle time results in
Figs.~\ref{fig:IT-1G-EPON},~\ref{fig:IT-1G-GPON},~\ref{fig:IT-10G-EPON},
and~\ref{fig:IT-10G-GPON}.

Considering online STP with excess grant sizing more closely, we
observe from Figs.~\ref{fig:OD-1G-EPON}a) and Fig.~\ref{fig:OD-10G-EPON}a)
that it achieves the smallest average packet delays for the
short $Z = 2$~ms maximum cycle duration.
Whereby, both online STP with excess grant sizing with reporting
at the end and with reporting at the beginning achieve similarly low
average delays, with reporting at the beginning giving only very minuscule
delay reductions for the mid-load range
of the $C = 1$~Gbps scenario in Fig.~\ref{fig:OD-1G-EPON}a).
Indeed, in additional evaluations that are not included in the plots
to avoid clutter, we have observed that STP with excess grant sizing
has similar average idle times as STP with limited grant sizing.
We observe for STP with limited grant sizing from
Fig.~\ref{fig:IT-1G-EPON}a) that
reporting at the beginning gives only very slight
idle time reductions in the mid-load range,
while both reporting approaches have essentially the
same idle times for the $C = 10$~Gbps scenario
in Fig.~\ref{fig:IT-10G-EPON}a).

\subsection{Double-phase Polling (D exc. shr.)}
Double-phase polling (DPP) with excess  sharing
has slightly higher delays and noticeably
longer idle times than online STP with excess
grant sizing throughout the scenarios considered in
Figs.~\ref{fig:OD-1G-EPON}--\ref{fig:IT-10G-GPON}. This is mainly
because DPP employs offline scheduling based on two ONU groups. That
is, the online polling processes to the two ONU groups are
interleaved, thus striving to mask the long $2 \tau$ idle period of
offline scheduling. This strategy is quite effective, as illustrated
by the dramatically lower packet delays and idle times compared to
the offline polling approaches. In fact, the average delays of DPP
approach quite closely those of online STP, but online STP achieves just a
little bit lower average delays mainly due to its more extensive
interleaving of the online polling processes to the individual ONUs.

The reporting strategy, reporting at the beginning or at the
end of the ONU transmission has essentially negligible impact on
both the average packet delays and the idle times.
This is mainly because the masking of idle times with the interleaving
of the two ONU polling groups is quite effective.
Further improving the interleaving by allowing an ONU group to
proceed with the scheduling earlier, i.e., after receiving
the last report message of the group at the beginning of the
last ONU transmission of the group versus the end of the last
ONU transmission has a very minor impact.

\subsection{Online MTP (M onl.)}
We observe from Fig.~\ref{fig:OD-1G-EPON}a) that for the short $Z=
2$~ms cycle length in the EPON, online MTP gives slightly higher
delays than online STP with excess allocation. In all other plots,
online MTP attains the smallest average packet delays. We observe
from Fig.~\ref{fig:OD-10G-EPON}b) and c) and
Fig.~\ref{fig:OD-10G-GPON} that for the higher speed $C = 10$~Gbps
and longer $Z = 4$ and 8~ms cycle lengths, online MTP achieves
slightly lower delays than online STP with excess allocation. We
also observe from these delay plots, as well as the idle time plots
in
Figs.~\ref{fig:IT-1G-EPON},~\ref{fig:IT-1G-GPON},~\ref{fig:IT-10G-EPON},
and~\ref{fig:IT-10G-GPON} that reporting at the beginning gives very
minor or no improvements compared to reporting at the end
in online MTP.

Online MTP exploits the interleaving of the polling processes to
the individual ONUs through the online scheduling framework as well as
the interleaving of multiple polling threads for each ONU.
Due to the multiple polling processes, i.e., more frequent polling,
the average upstream transmission window lengths with MTP
are typically smaller than with STP~\cite{MeMR13,Muk01}.
Shifting the reporting from the end to the beginning of
an upstream transmission constitutes therefore a smaller
shift of the report message compared to STP with its longer
transmission windows.
In addition, the multiple levels of interleaving in online MTP
leave little unmasked idle times that could be shortened
by shifting the report message to the beginning.

\section{Conclusion}
\label{concl:sec}
We have examined the effects of report message scheduling, specifically,
scheduling the report message at the beginning or at the end of the
upstream transmission of a optical network unit (ONU) in
a passive optical network (PON).
We have examined these two extreme positions of the report message
(beginning or end of the upstream transmission) for a wide range
of dynamic bandwidth allocation (DBA) mechanisms in an
Ethernet PON (EPON) and Gigabit PON (GPON) for both 1~Gbps and 10~Gbps
upstream channel bandwidth.
Aside from providing insights into the effects of report message scheduling,
this study provides insights into the performance of a wide range of
DBA approaches at the 10~Gbps channel bandwidth for long-reach PONs (LRPONs).
Most prior studies have only considered the 1~Gbps bandwidth.

We have found that report scheduling at the beginning
achieves significant reductions of channel idle time and average
packet delays for DBAs with the offline scheduling framework that
requires reports from all ONUs before sizing and scheduling the
upstream transmission windows for the next polling cycle.
This is accomplished by reducing the unmasked idle time period, which
is one round-trip propagation delay $2 \tau$ for reporting at the end,
by the duration of the payload transmission
time of one ONU by reporting at the beginning.

DBA approaches with short or few unmasked idle times provide
little opportunity for increasing the masking of idle time
through shifting the position of the report message.
Thus, we observed that online single-thread polling (STP) that interleaves
polling processes to the individual ONUs,
double-phase polling (DPP)~\cite{Sam01} that interleaves offline
polling processes to two ONU groups, as well as online multi-thread polling
(MTP)~\cite{MeMR13} are largely insensitive to the
report scheduling.

There are several important direction for future research on
effective dynamic bandwidth allocation for
PON access networks.
One direction is to integrated the PON DBA mechanisms
with access networks involving other transmission
media~\cite{KrDR12,Luo13}, such as
wireless
networks~\cite{AuLMR14,Bur04,Kun01,MiTK12,RaWL12}.
Another direction is to streamline the internetworking
of access networks with metro
area networks~\cite{BiBC13,MaR04,MaRW03,ScMRW03,YaMRC03,YuCL10}
and then with wide area networks, through
specific network integration and internetworking mechanisms.

\vspace{\baselineskip}

\noindent
{\textsc {Appendix~I: Analysis of Channel Idle Time}}

In this appendix,
we build on the idle time analysis for reporting at the end
for single-thread polling and DPP in~\cite{Mar01} as well as
the analysis for reporting at the end for multi-thread polling
in~\cite{MeMR13} to analyze the idle time for both single- and
multi-thread polling with reporting at the beginning.
We then analyze the reduction of the idle achieved
by reporting at the beginning.

Note from (\ref{I:eqn}) that the channel idle time is the difference
between the instant $\alpha(n, \theta, j)$ when the beginning of the
upstream transmission of ONU $j$ of thread $\theta$ in cycle $n$
starts to arrive at the OLT and the instant $\Omega(n, \theta, j)$
when the end of the preceding ONU transmission arrives at the OLT.
We first determine $\Omega(n, \theta, j)$ for the various
combinations of scheduling frameworks and ONU indices, as summarized
in Table~\ref{Om:tab}.
\begin{table}
\caption{Time instant $\Omega(n, \theta, j)$ of end of arrival
of upstream transmission preceding the arrival of upstream
transmission of ONU $j$ of thread $\theta$ in cycle $n$ at OLT.}
\label{Om:tab}
\begin{tabular}{|l|l|l|} \hline
Scheduling   &   Thread and    &                                \\
framework    &   ONU indices   & $\Omega(n, \theta, j) = $      \\
\hline
 && \\
STP (both offl.  &   $j = 1$           &   $\beta(n-1, O)$          \\
and online)       &   $2 \leq j \leq O$  & $\beta(n, j-1)$        \\ \hline
&& \\
MTP     &      $\theta = 1; j = 1 $     & $\beta(n-1,\Theta, O) $   \\
        &  $2 \leq \theta \leq \Theta;\ j = 1 $ & $\beta(n, \theta-1, O)$ \\
   & $1 \leq \theta \leq \Theta;\ 2 \leq j \leq O$ & $\beta(n, \theta, j-1)$ \\
\hline
\end{tabular}
\end{table}
For single-thread polling,
both with online and offline scheduling, the transmission
of the last ONU $j = O$ of the preceding cycle $n-1$ precedes
the arrival of the transmission of the first ONU $j = 1$ in cycle $n$,
i.e., $\Omega(n, j = 1) = \beta(n-1, O)$.
In turn, the arrival of the transmission of ONU $j = 1$
in cycle $n$ precedes the arrival of the transmission of
ONU $j = 2$; in general for the ONUs ``within'' a single-thread polling
cycle, the arrival of the transmission from ONU $j-1$ precedes
the arrival of the transmission from ONU $j,\ j = 2, 3, \ldots, O$.
For multi-thread polling, the transmission of the last ONU $j = O$
in the last thread $\theta = \Theta$ of a cycle $n-1$
precedes the arrival from the first ONU $j = 1$ of the first thread
$\theta = 1$ of the subsequent cycle $n$.
The first transmission of each subsequent thread $\theta = 2, \ldots, \Theta$
within cycle $n$, is preceded by
the last transmission $j = O$ of the preceding thread $\theta-1$.
The second and subsequent transmissions $j = 2, 3, \ldots, O$
within a thread are preceded by the preceding ONU transmission $j-1$.

As outlined in Section~\ref{Isum:sec}, the idle time constraint and
the signaling constraint determine the arrival time instant
$\alpha(n, \theta, j)$ of ONU transmission $j$ of thread $\theta$ in
cycle $n$ at the OLT. Specifically, the idle time constraint
requires that the arrival instant $\alpha(n, \theta, j)$ is no
earlier than a guard time $t_g$ after the end of the arrival of the
preceding transmission at instant $\Omega(n, \theta, j)$. The
signaling constraint imposes the gate signaling delay $T(n, \theta,
j)$ between the scheduling instant $\gamma(n, \theta, j)$ of the ONU
transmission and its arrival at the OLT. Thus,
\begin{eqnarray} \label{al1:eqn}
\alpha(n, \theta, j) = \max \{ \Omega(n, \theta, j) + t_g,\
                         \gamma(n, \theta, j) + T(n, \theta, j) \}.
\end{eqnarray}
Inserting the expression (\ref{al1:eqn}) for $\alpha(n, \theta, j)$
into equation (\ref{I:eqn}) for evaluating the channel idle time gives
\begin{eqnarray} \label{I2:eqn}
I(n, \theta, j) = \max \{ t_g,\ \gamma(n, \theta, j)
             + T(n, \theta, j)  - \Omega(n, \theta, j) \}.
\end{eqnarray}

We evaluate the reduction of the channel idle time with
report scheduling at the beginning compared to reporting at the end
as
\begin{eqnarray}
\Delta_I \! &= & \! I_{\beta}(n, \theta, j) - I_{\alpha}(n, \theta, j) \\
     \! & = & \! \max \{ t_g,\ \gamma_{\beta}(n, \theta, j)  \label{DI:eqn}
             + T(n, \theta, j)  - \Omega(n, \theta, j) \}  \\
     && \ \ - \max \{ t_g,\ \gamma_{\alpha}(n, \theta, j)
             + T(n, \theta, j)  - \Omega(n, \theta, j) \}. \nonumber
\end{eqnarray}
For the further analysis of (\ref{DI:eqn}), note that the
scheduling instant with reporting at the end $\gamma_{\beta}(n, \theta, j)$ is
always later (or at the same time) than the scheduling instant with reporting
at the beginning $\gamma_{\alpha}(n, \theta, j)$.
Specifically, these two time instants are the same, when the
corresponding upstream transmission carries only the report message
and no payload data.
If the upstream transmission carries some payload data, then
these two time instants are separated by the transmission time
for the carried payload data.
Thus,
\begin{eqnarray} \label{gabeal:eqn}
\gamma_{\beta}(n, \theta, j) \geq \gamma_{\alpha}(n, \theta, j).
\end{eqnarray}

\begin{table}
\caption{Summary of cases for reduction $\Delta_I$ of channel idle time
with report scheduling at the beginning compared to report scheduling
at the end of an ONU upstream transmission}
\label{DI:tab}
\begin{tabular}{|l|l|} \hline
Case   &    $\Delta_I = $  \\ \hline
 & \\
$\gamma_{\beta}(n, \theta, j)
   + T(n, \theta, j)  - \Omega(n, \theta, j) \leq t_g$  &
                                                      $0$   \\ \hline
 & \\
$\gamma_{\beta}(n, \theta, j)
   + T(n, \theta, j)  - \Omega(n, \theta, j) > t_g$   &
           $\gamma_{\beta}(n, \theta, j) + T(n, \theta, j)$     \\
\ \ $\gamma_{\alpha}(n, \theta, j)
   + T(n, \theta, j)  - \Omega(n, \theta, j) < t_g$
                   & \ \  $- \Omega(n, \theta, j) - t_g$ \\
               & \ \ \ \ \ \ \ \ \ \ \ \ \ \ \ $< \gamma_{\beta} - \gamma_{\alpha}$ \\ \hline
& \\
$\gamma_{\alpha}(n, \theta, j)
   + T(n, \theta, j)  - \Omega(n, \theta, j) \geq t_g$  &
                                  $\gamma_{\beta} - \gamma_{\alpha}$ \\
\hline
\end{tabular}
\end{table}
Considering (\ref{DI:eqn}), there are three cases for evaluating
the channel idle time.
First, in case
\begin{eqnarray}
\gamma_{\beta}(n, \theta, j)
   + T(n, \theta, j)  - \Omega(n, \theta, j) \leq t_g,
\end{eqnarray}
(\ref{gabeal:eqn}) implies that also
\begin{eqnarray}
\gamma_{\alpha}(n, \theta, j)
   + T(n, \theta, j)  - \Omega(n, \theta, j) \leq t_g.
\end{eqnarray}
Thus, both maxima in (\ref{DI:eqn}) are attained by $t_g$
and the resulting reduction in channel idle time is zero,
as summarized in Table~\ref{DI:tab}.
Intuitively, this first case occurs if the preceding ONU
transmission, of which the end arrives to the OLT at $\Omega(n, \theta, n)$
is sufficiently long to mask the signaling time for the
transmission of ONU $j$ of thread $\theta$ in cycle $n$.

The other extreme case is that
\begin{eqnarray}
\gamma_{\alpha}(n, \theta, j)
   + T(n, \theta, j)  - \Omega(n, \theta, j) \geq t_g,
\end{eqnarray}
which implies by (\ref{gabeal:eqn})
that also
\begin{eqnarray}
\gamma_{\beta}(n, \theta, j)
   + T(n, \theta, j)  - \Omega(n, \theta, j) \geq t_g.
\end{eqnarray}
Hence, both maxima in (\ref{DI:eqn}) are attained by the terms
involving the scheduling instants $\gamma$; specifically, $\Delta_I
= \gamma_{\beta} - \gamma_{\alpha}$. That is, the reduction in
channel idle time is equal to the duration of the transmission time
of the payload of the ONU transmission. This case occurs if the
signaling delay for ONU transmission $j$ of thread $\theta$ in cycle
$n$ is not masked by the preceding ONU transmission. Such an
unmasked idle time can occur if the preceding ONU transmission is
too short to mask the gate signalling delay. Or the polling
structure introduces a mandatory idle time that cannot be masked by
a preceding transmission. For instance, offline scheduling requires
the receipt of the report from the last ONU transmission ($j = O$)
in cycle $n-1$ before sizing and scheduling the grants for cycle
$n$. That is, the first ONU transmission ($j=1$) in a cycle $n$, is
preceded by the last ONU transmission of the preceding cycle and
consequently, the time instant of the end of the preceding ONU
transmission is $\Omega(n, 1) = \beta(n-1, O)$, see
Table~\ref{Om:tab}, which coincides with the scheduling instant
$\gamma(n, 1)$, see Table~\ref{gam:tab} for reporting at the end.
The resulting idle period (\ref{I2:eqn}) is the gate signalling
delay $T(n, 1)$ which equals one gate transmission time $t_G$ (typ.
negligible) and the round-trip time $2 \tau$. This idle time can be
reduced through shifting the report message to the beginning.
Specifically, the unmasked idle time can be reduced by $\Delta_I =
\gamma_{\beta} - \gamma_{\alpha}$, i.e., the transmission time for
the payload in the last ONU ($j = O$) transmission in cycle $n-1$.

The intermediate case is that
\begin{eqnarray}
\gamma_{\beta}(n, \theta, j)
   + T(n, \theta, j)  - \Omega(n, \theta, j) > t_g,
\end{eqnarray}
while
\begin{eqnarray} \label{case3al:eqn}
\gamma_{\alpha}(n, \theta, j)
   + T(n, \theta, j)  - \Omega(n, \theta, j) < t_g.
\end{eqnarray}
In this case, the maximum in the first line of (\ref{DI:eqn})
is attained by the term involving $\gamma_{\beta}$, while the maximum
in the second line is attained by $t_g$.
Thus,
\begin{eqnarray}
\Delta_I = \gamma_{\beta}(n, \theta, j)
   + T(n, \theta, j)  - \Omega(n, \theta, j)  - t_g,
\end{eqnarray}
which by (\ref{case3al:eqn}) is less than $\gamma_{\beta} - \gamma_{\alpha}$.

\vspace{\baselineskip}
\noindent
{\textsc {Appendix~II: Evaluation of Optimized Report Scheduling}}
\begin{figure}[t!]
\centering
\begin{tabular}{c}
\includegraphics[scale=0.65]{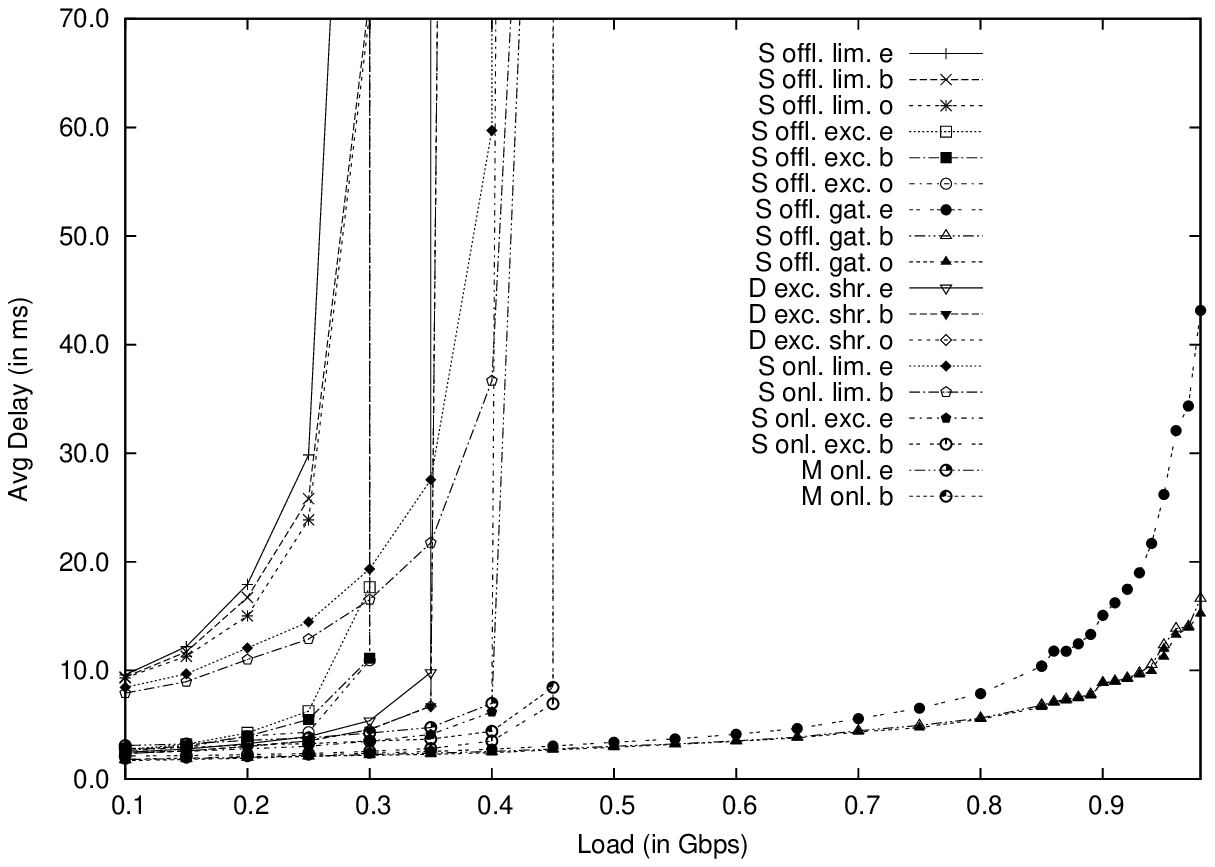} \\
\footnotesize{a) Max. cycle length $Z=2$~ms} \\
\includegraphics[scale=0.65]{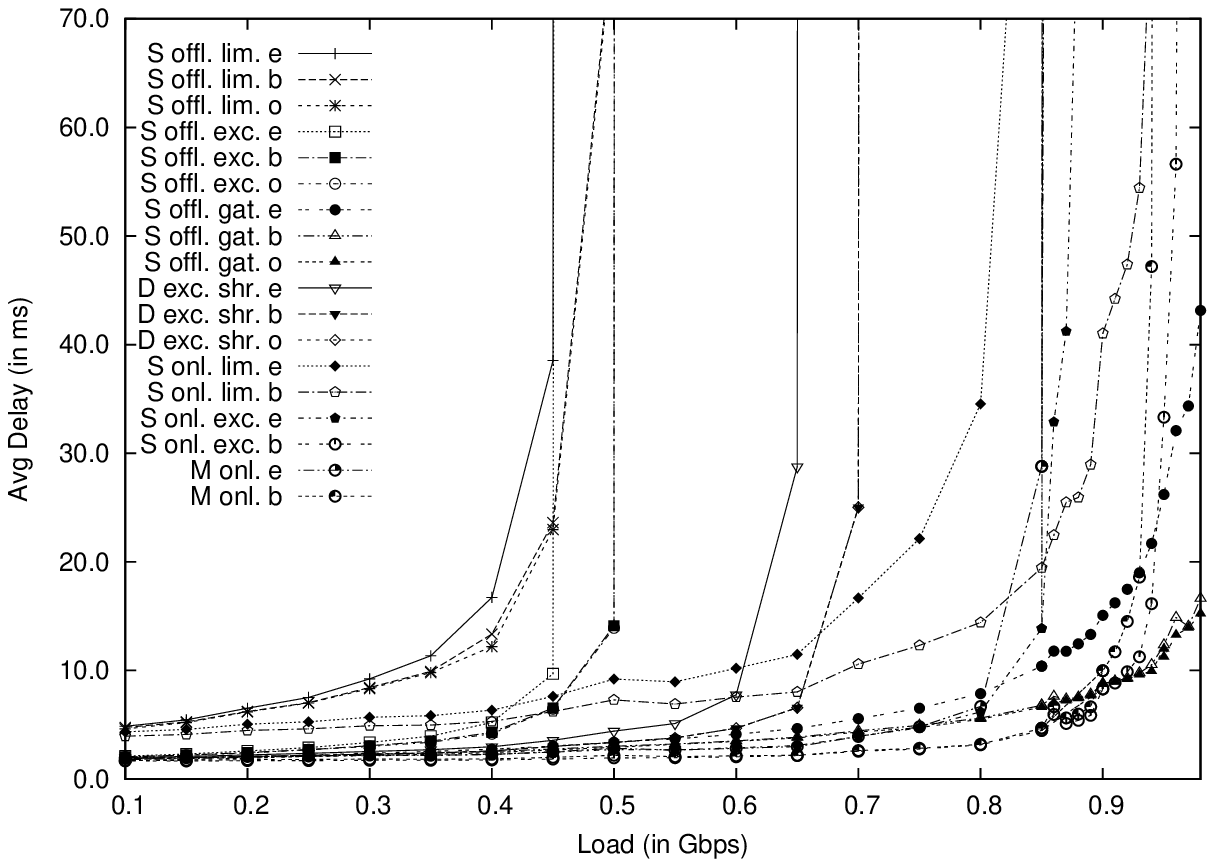} \\
\footnotesize{b)  Max. cycle length $Z=4$~ms} \\
\includegraphics[scale=0.65]{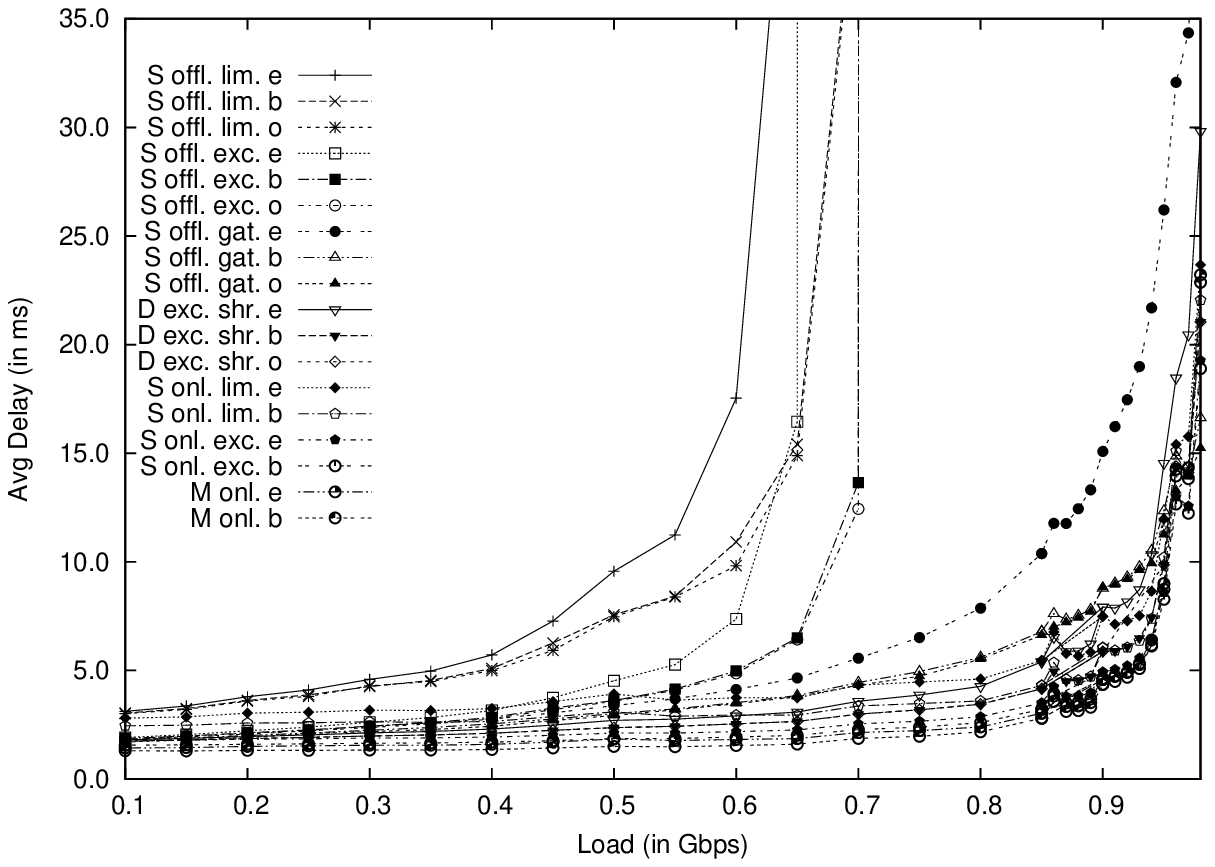} \\
\footnotesize{c)  Max. cycle length $Z=8$~ms}\\
\end{tabular}
\caption{Mean packet delay for EPON with upstream bandwidth
      $C = 1$~Gbps, $O=8$ ONUs, and 100~km range.}
\label{fig:OD-1G-8o}
\end{figure}
\begin{figure}[t!]
\centering
\begin{tabular}{c}
\includegraphics[scale=0.65]{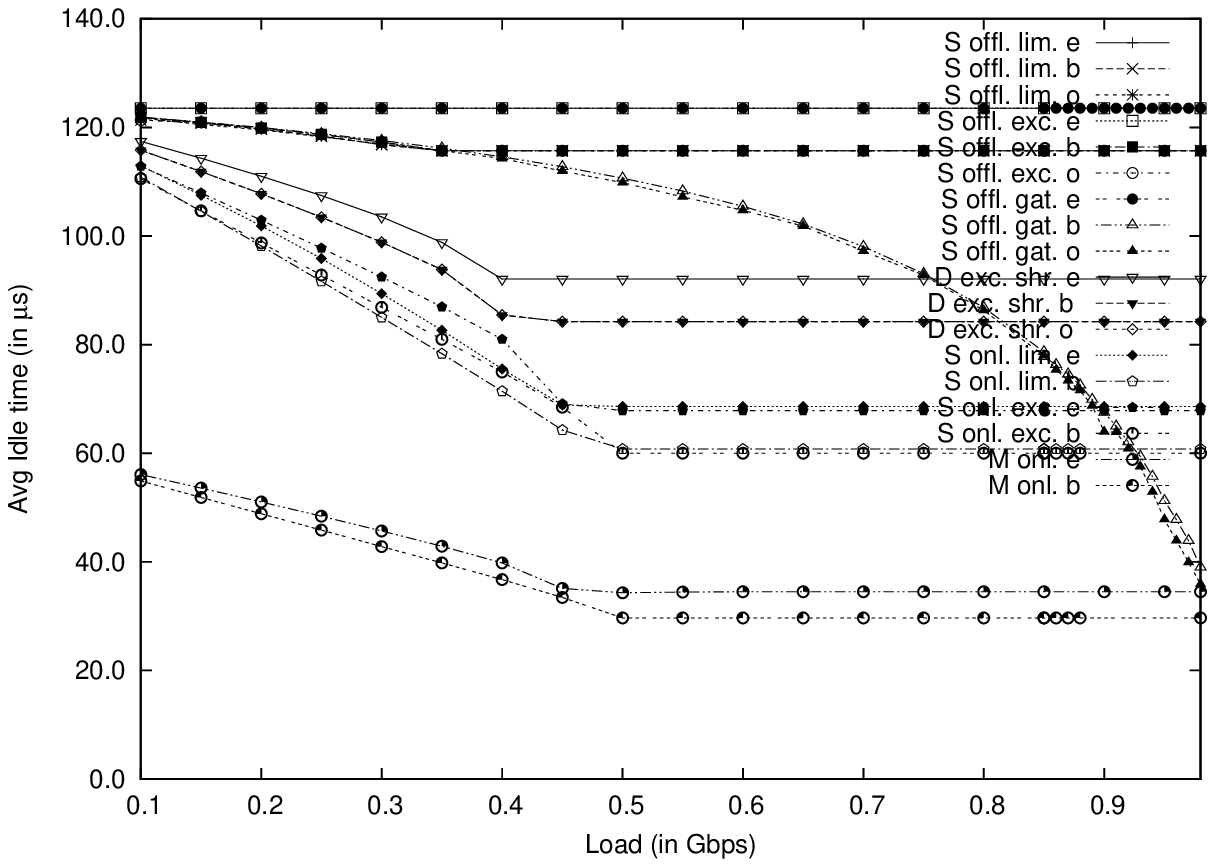} \\
\footnotesize{a) Max. cycle length $Z=2$~ms} \\
\includegraphics[scale=0.65]{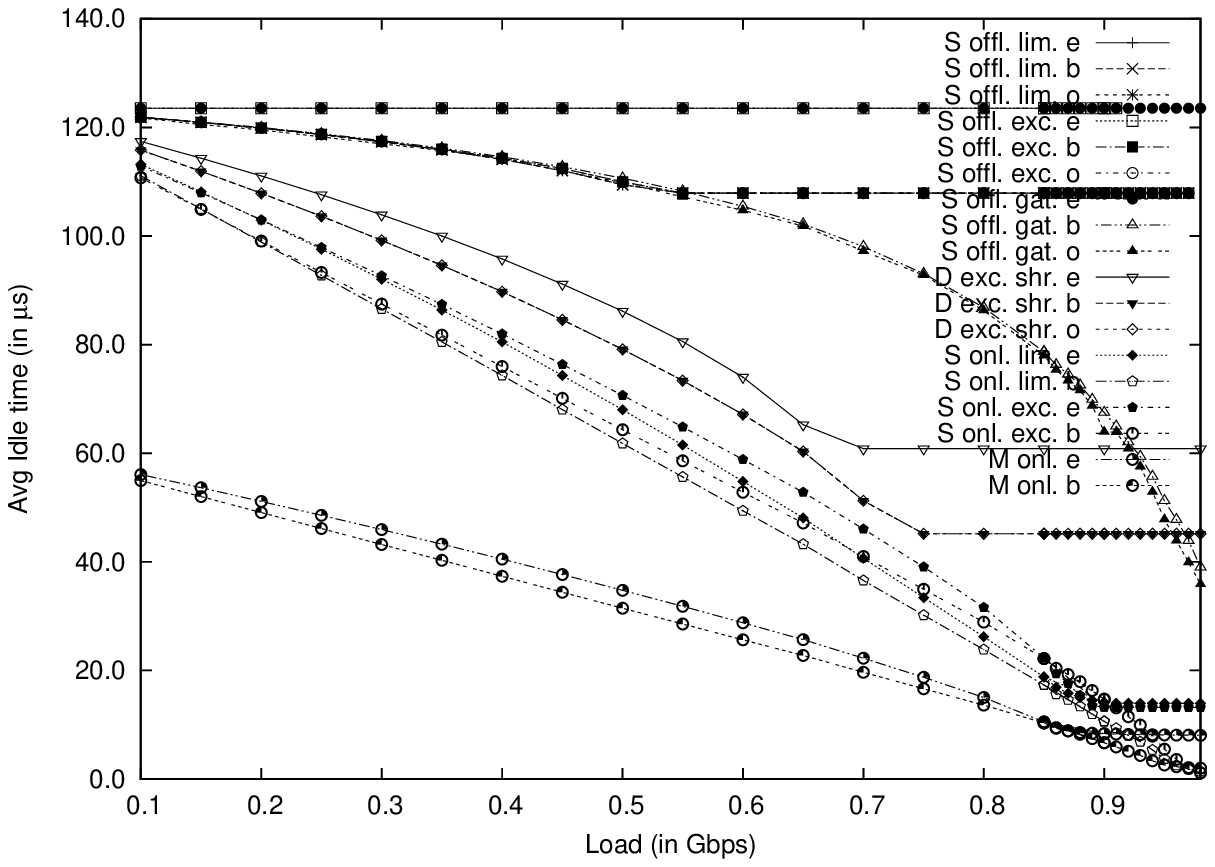} \\
\footnotesize{b)  Max. cycle length $Z=4$~ms} \\
\includegraphics[scale=0.65]{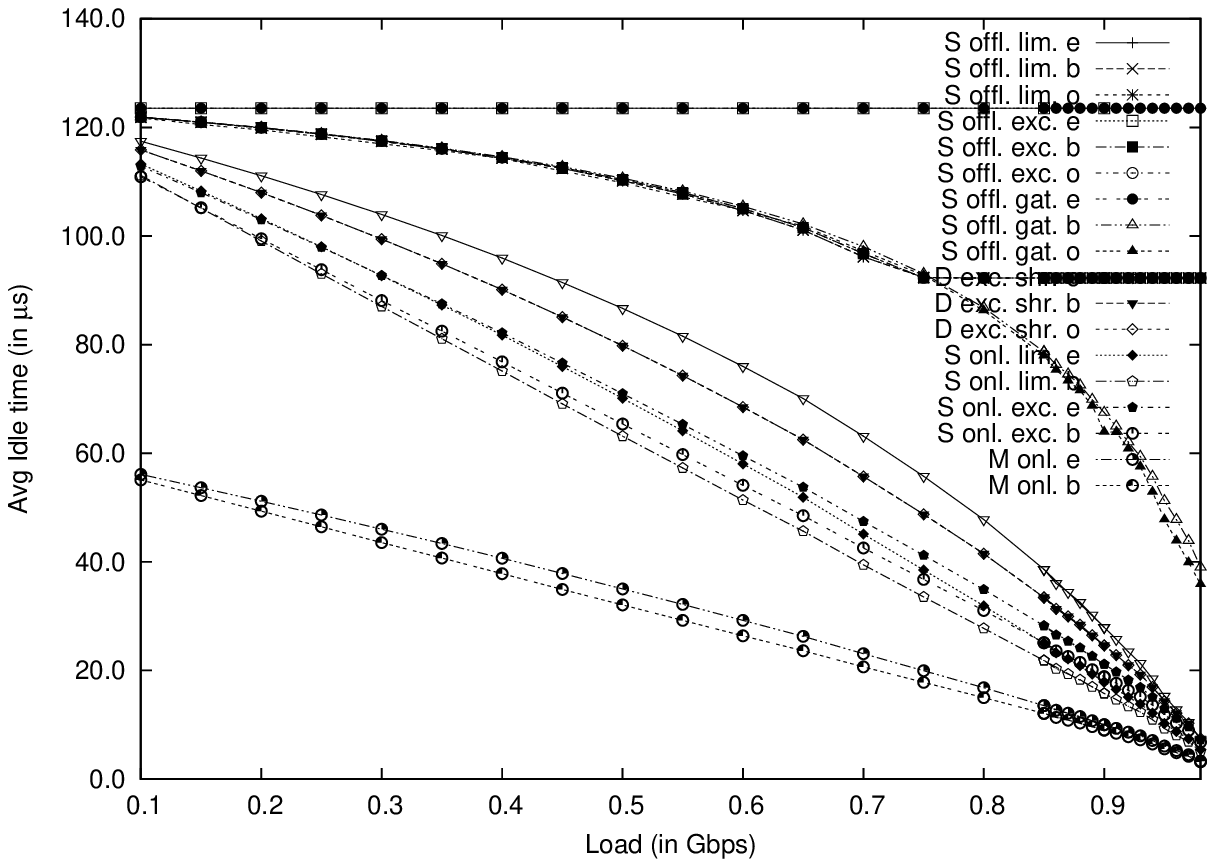} \\
\footnotesize{c)  Max. cycle length $Z=8$~ms}\\
\end{tabular}
\caption{Mean idle time for EPON with upstream bandwidth
   $C = 1$~Gbps, $O=8$ ONUs, and 100~km range.}
\label{fig:IT-1G-8o}
\end{figure}

\begin{figure}[t!]
\centering
\begin{tabular}{c}
\includegraphics[scale=0.65]{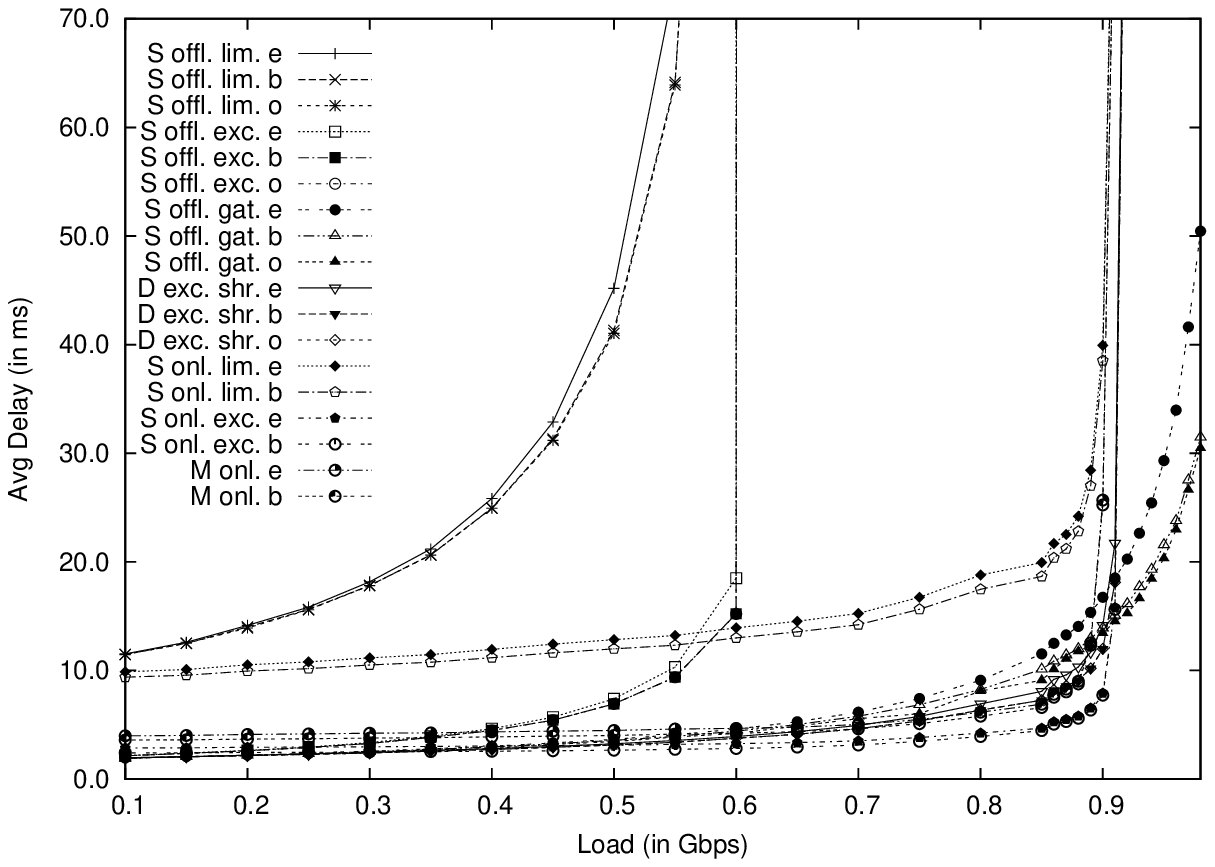} \\
\footnotesize{a) Max. cycle length $Z=2$~ms} \\
\includegraphics[scale=0.65]{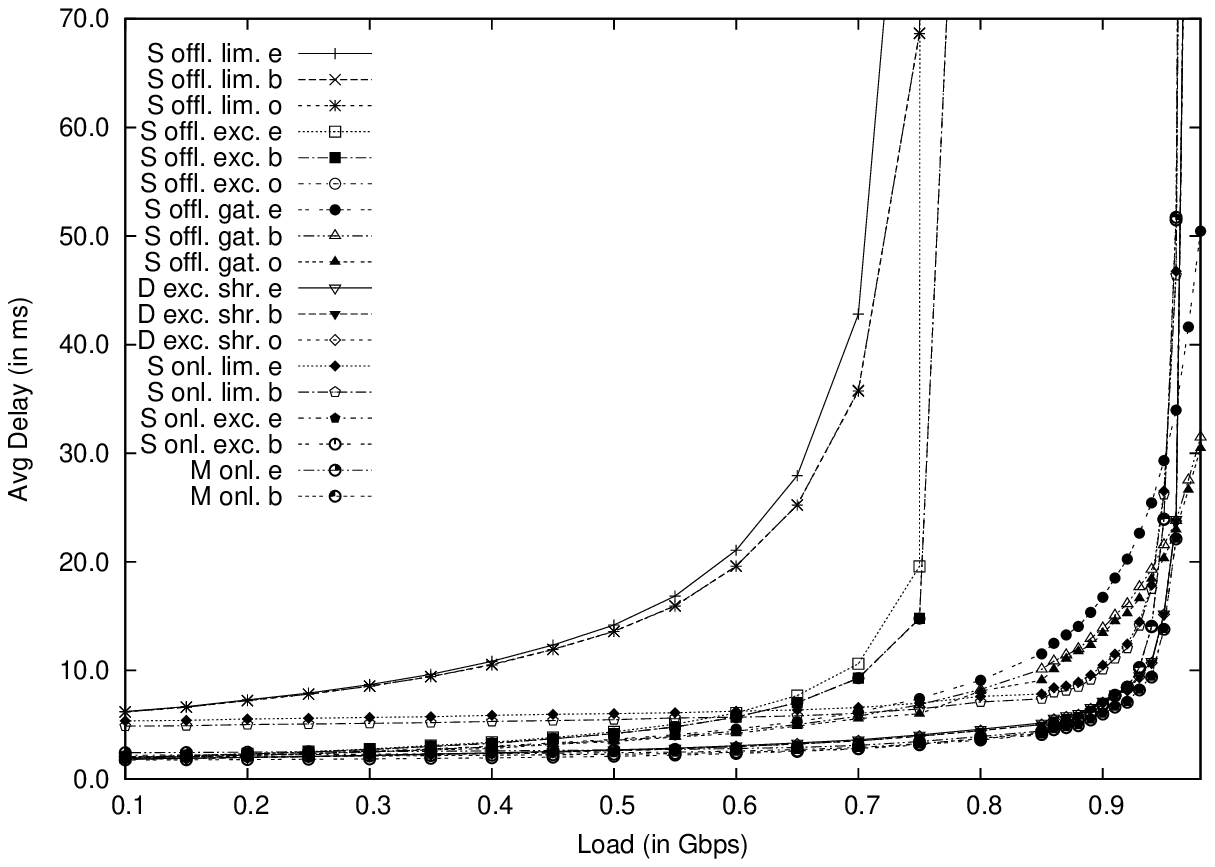} \\
\footnotesize{b)  Max. cycle length $Z=4$~ms} \\
\includegraphics[scale=0.65]{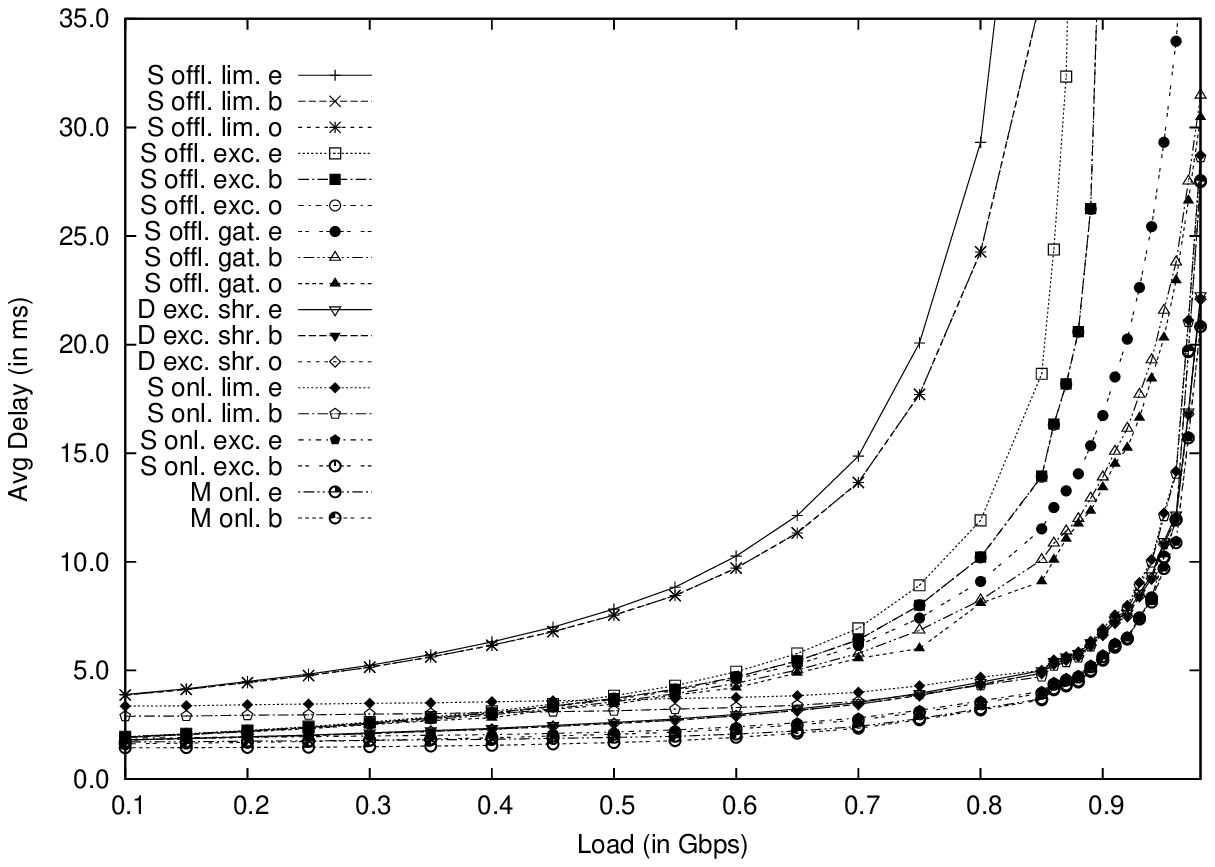} \\
\footnotesize{c)  Max. cycle length $Z=8$~ms}\\
\end{tabular}
\caption{Mean packet delay for EPON with upstream bandwidth
$C = 1$~Gbps, $O=32$ ONUs, and 100~km range.}
\label{fig:OD-1G-32o}
\end{figure}
\begin{figure}[t!]
\centering
\begin{tabular}{c}
\includegraphics[scale=0.65]{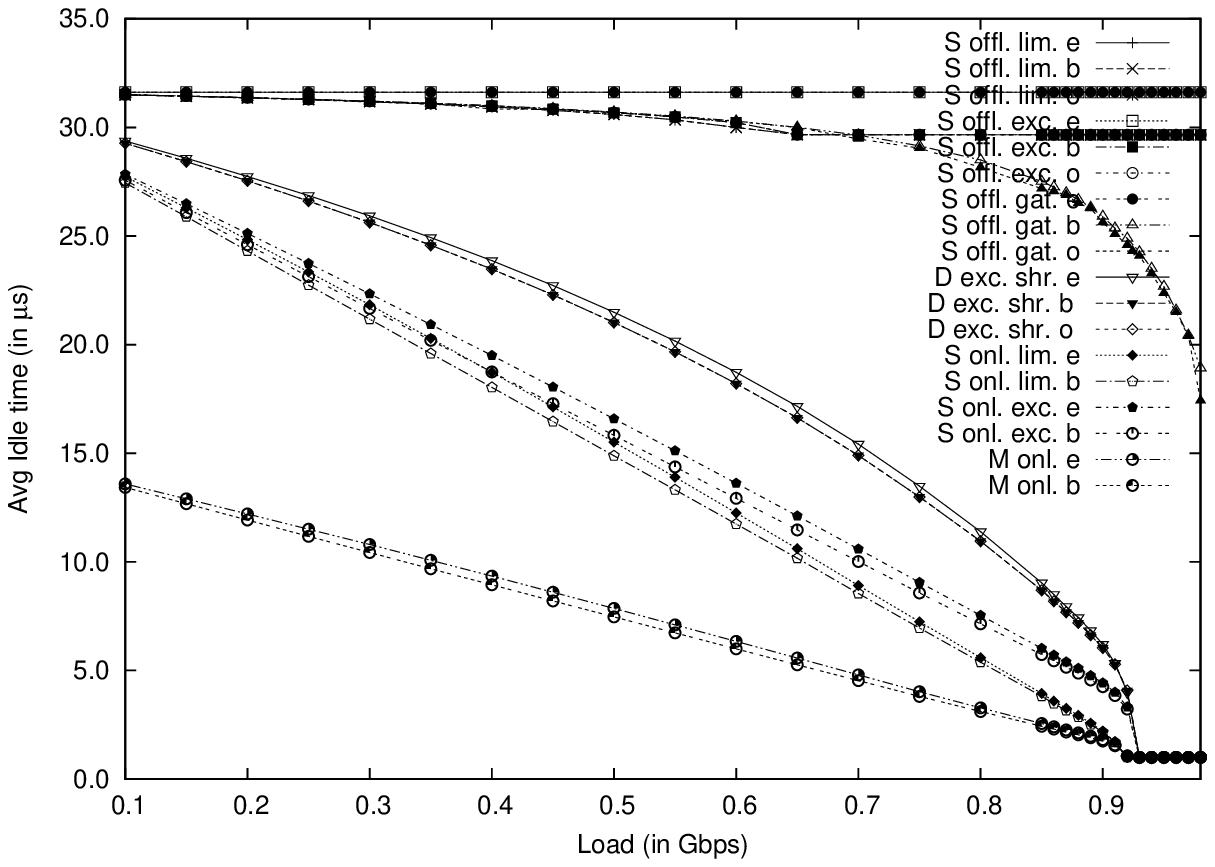} \\
\footnotesize{a) Max. cycle length $Z=2$~ms} \\
\includegraphics[scale=0.65]{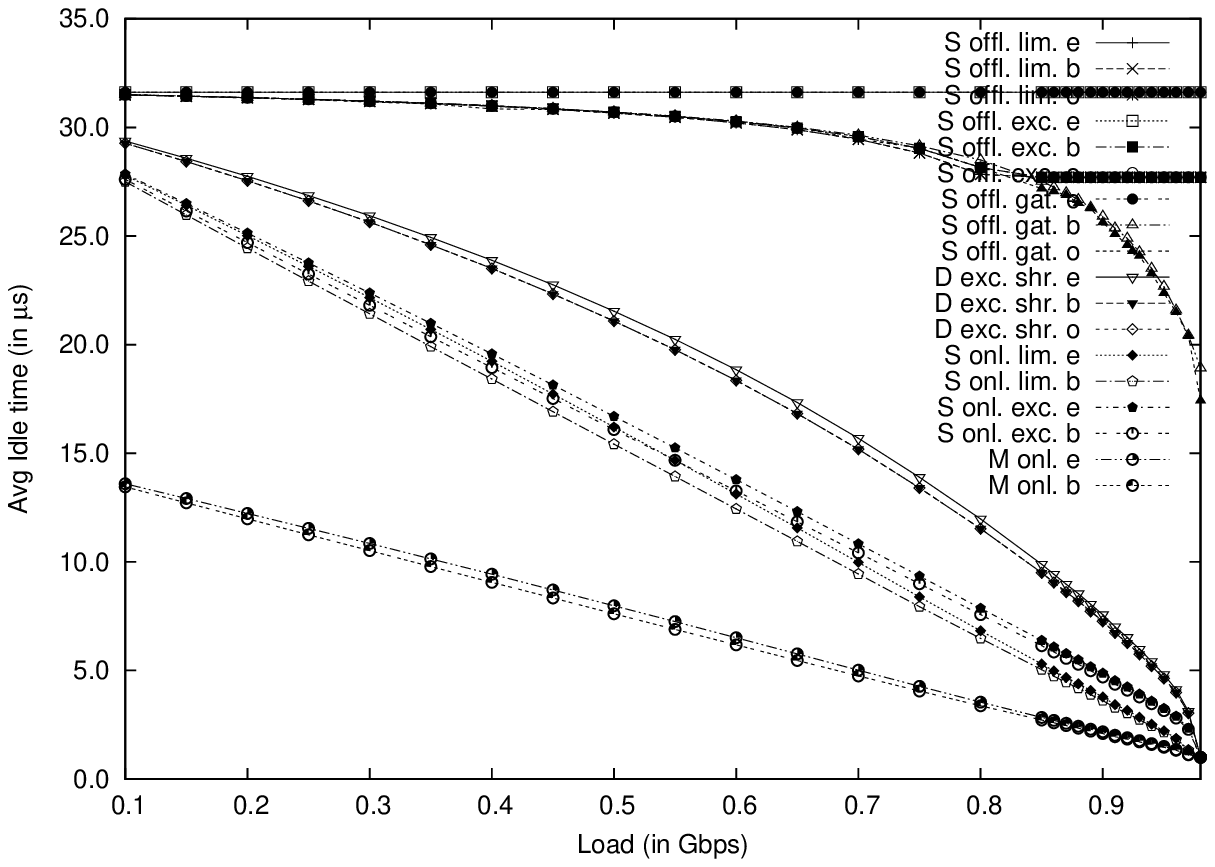} \\
\footnotesize{b)  Max. cycle length $Z=4$~ms} \\
\includegraphics[scale=0.65]{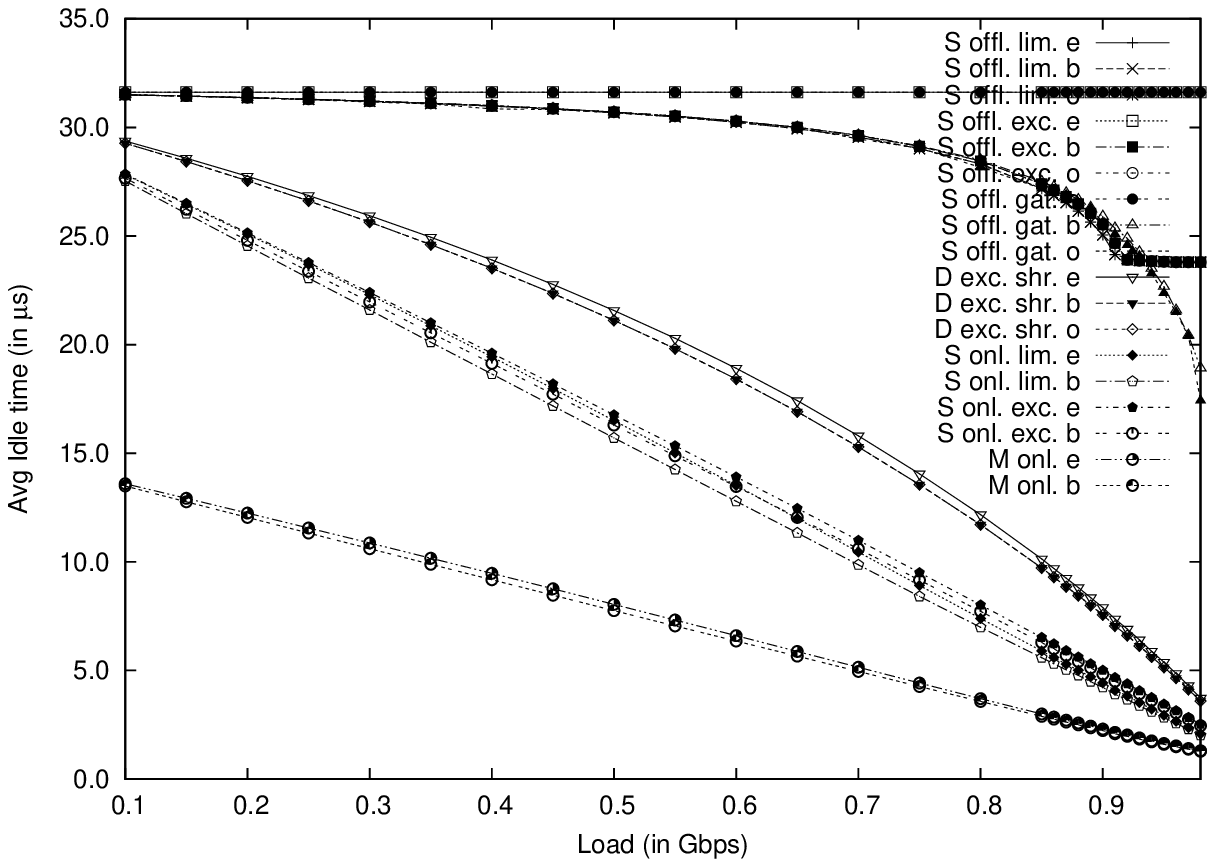} \\
\footnotesize{c)  Max. cycle length $Z=8$~ms}\\
\end{tabular}
\caption{Mean idle time for EPON with upstream bandwidth
     $C = 1$~Gbps, $O=32$ ONUs, and 100~km range.}
\label{fig:IT-1G-32o}
\end{figure}

\begin{figure}[t!]
\centering
\begin{tabular}{c}
\includegraphics[scale=0.65]{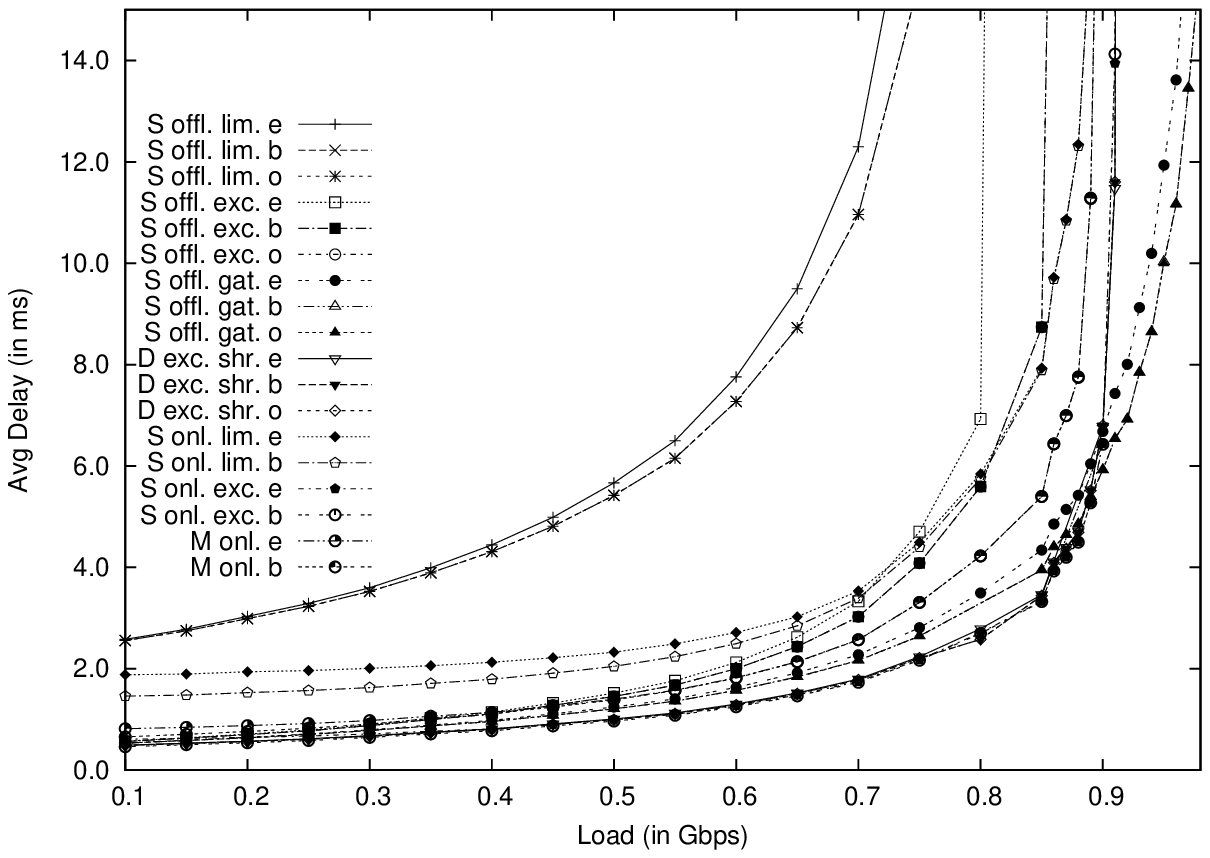} \\
\footnotesize{a) Max. cycle length $Z=2$~ms} \\
\includegraphics[scale=0.65]{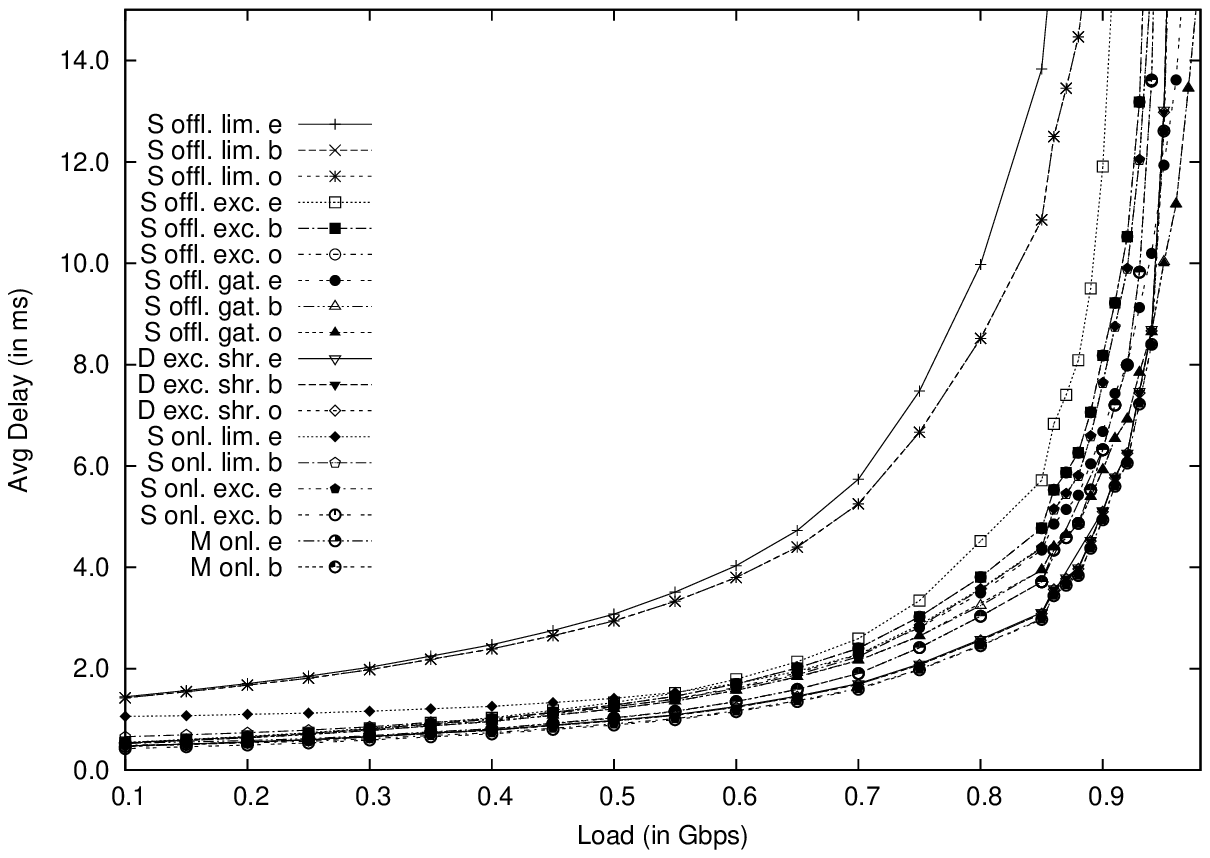} \\
\footnotesize{b)  Max. cycle length $Z=4$~ms} \\
\includegraphics[scale=0.65]{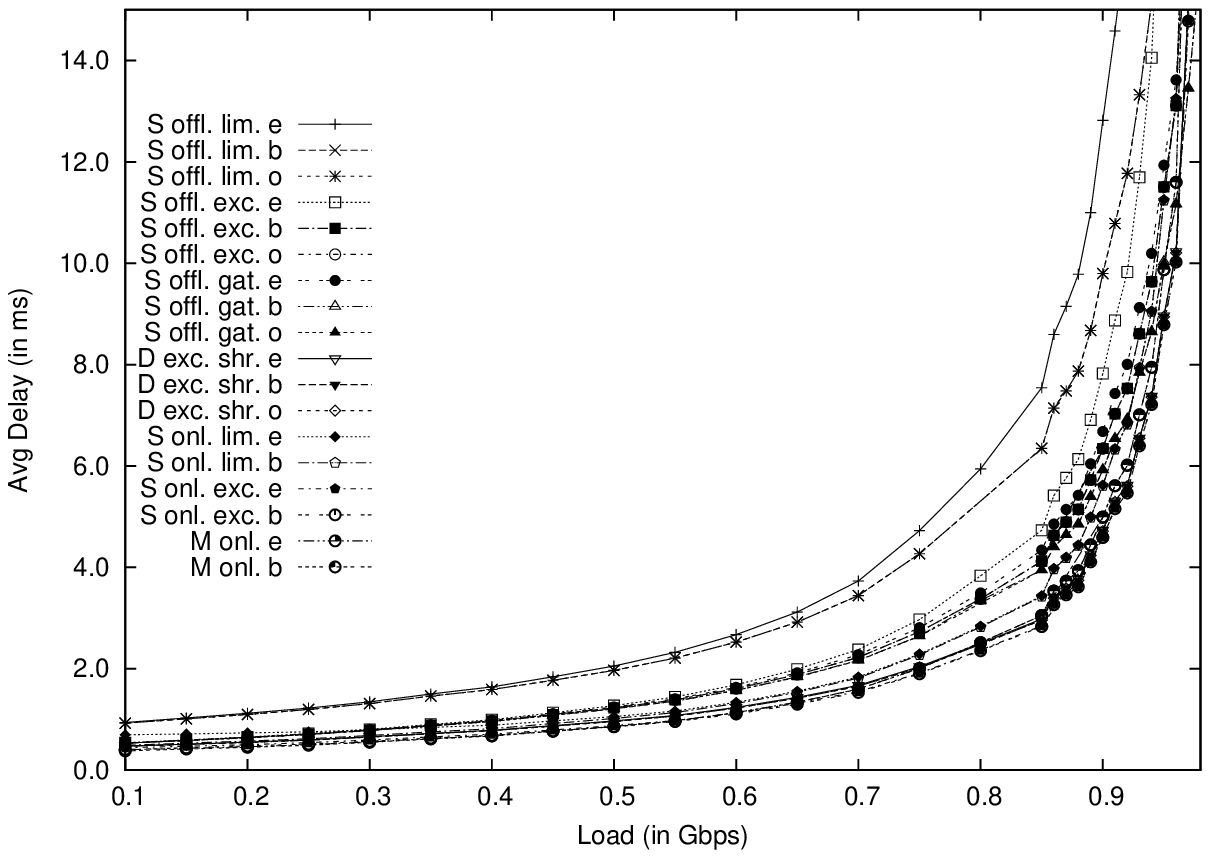} \\
\footnotesize{c)  Max. cycle length $Z=8$~ms}\\
\end{tabular}
\caption{Mean packet delay for EPON with upstream bandwidth
  $C = 1$~Gbps, $O=32$ ONUs, and $20$~km range.}
\label{fig:OD-1G-32o_20}
\end{figure}
\begin{figure}[t!]
\centering
\begin{tabular}{c}
\includegraphics[scale=0.65]{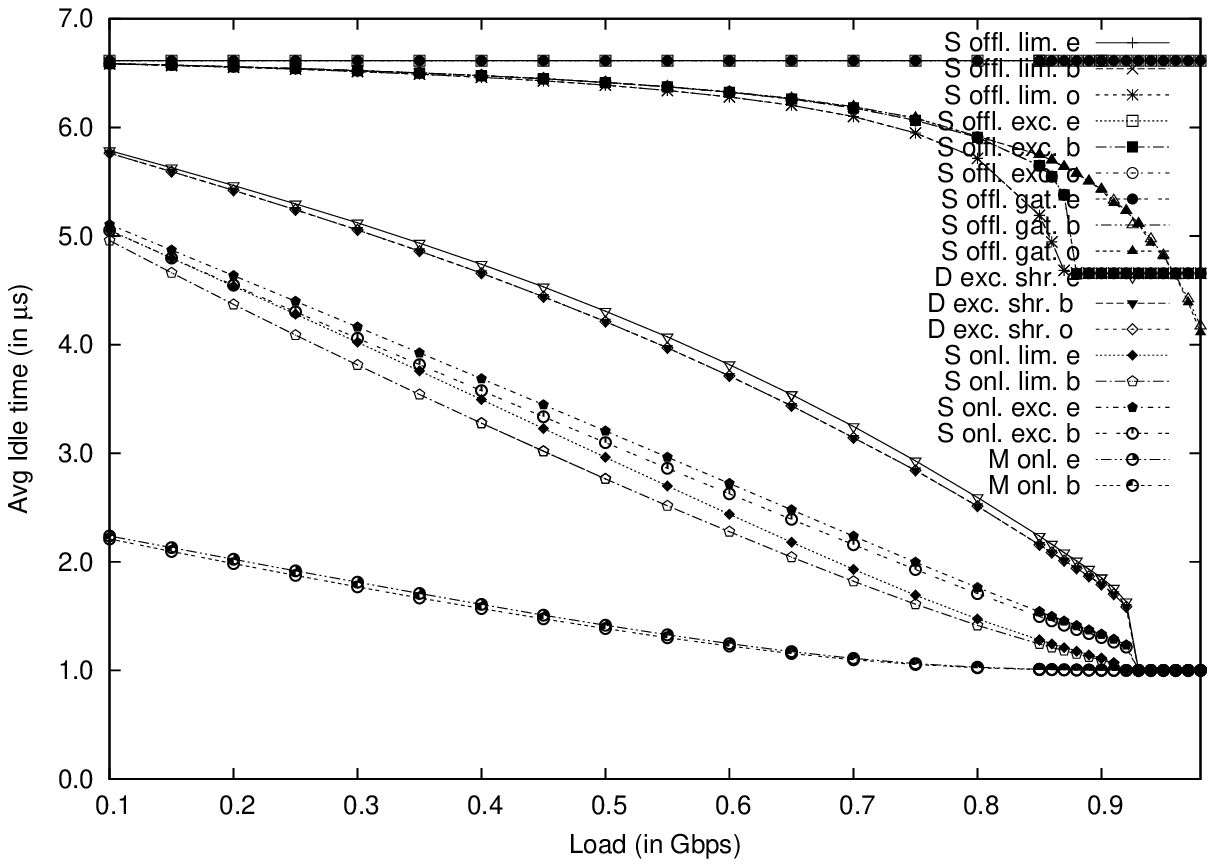} \\
\footnotesize{a) Max. cycle length $Z=2$~ms} \\
\includegraphics[scale=0.65]{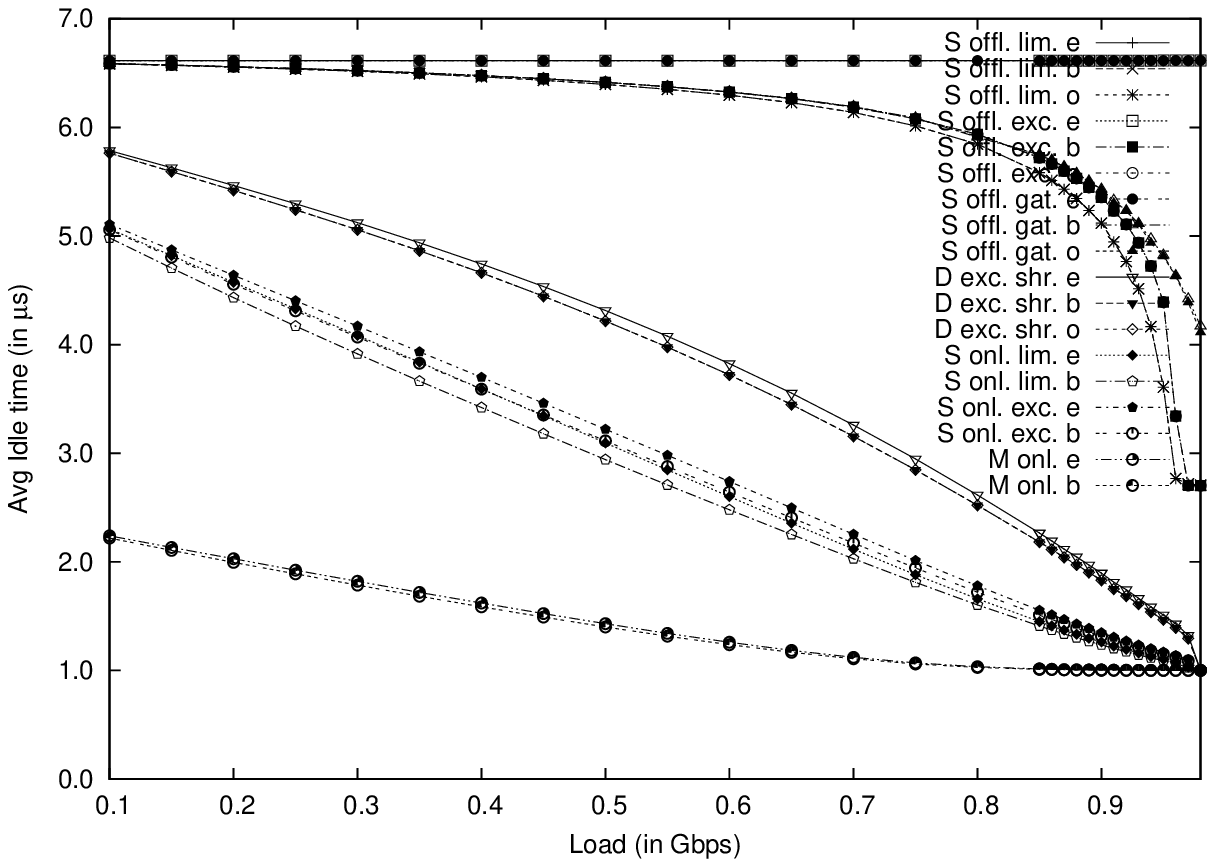} \\
\footnotesize{b)  Max. cycle length $Z=4$~ms} \\
\includegraphics[scale=0.65]{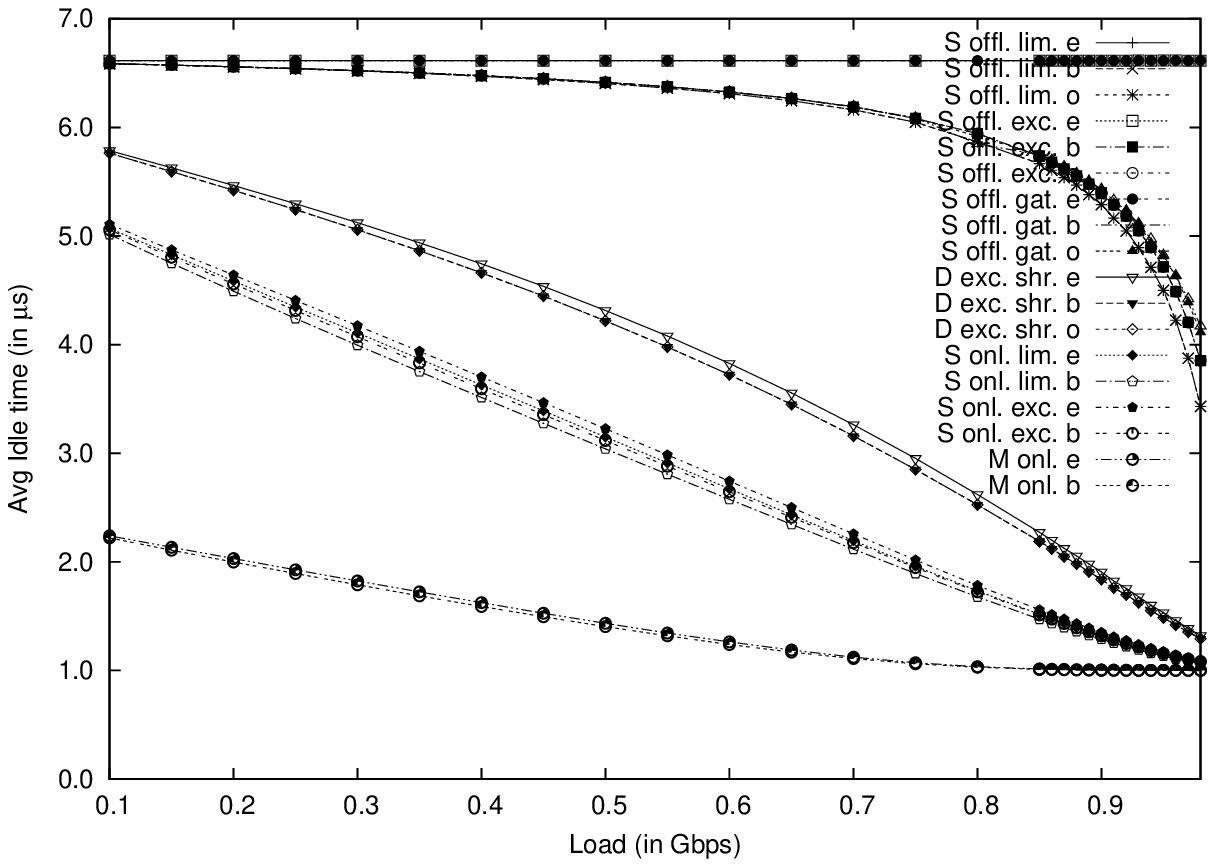} \\
\footnotesize{c)  Max. cycle length $Z=8$~ms}\\
\end{tabular}
\caption{Mean idle time for EPON with upstream bandwidth
     $C = 1$~Gbps, $O=32$ ONUs, and $20$~km.}
\label{fig:IT-1G-32o_20}
\end{figure}
In Figs.~\ref{fig:OD-1G-8o}--\ref{fig:IT-1G-32o} we plot the mean
delays and idle times for all DBA mechanisms considered in this
study, see Section~\ref{sch_beg_end:sec}. For the DBA approaches
with offline scheduling, including DPP, we plot results for the
three examined report scheduling approaches, namely reporting at the
end (e), reporting at the beginning (b), and the dynamically
optimized report message scheduling (o) of Section~\ref{opt:sec}. We
consider an $C = 1$~Gbps EPON with $O$ ONUs placed randomly at a
distance between 90 and 100~km from the OLT.

We observe from Figs.~\ref{fig:OD-1G-8o}--\ref{fig:IT-1G-32o} that
the DBA approaches with offline scheduling follow the general
pattern of results from Table~\ref{O:tab} for the different report
message scheduling approaches. That is, for a small number of $O =
8$ ONUs, see Figs.~\ref{fig:OD-1G-8o} and~\ref{fig:IT-1G-8o},
reporting at the beginning achieves significantly lower mean packet
delays and lower mean idle times than reporting at the end.
Optimized report message scheduling, in turn, achieves somewhat
lower packet delays and idle times than reporting at the beginning
for offline limited and offline gated grant sizing. For offline
excess and DPP, which are more efficient due to better grant sizing
and masking of idle times, there are no visible improvements with
optimized report message scheduling. We observe from
Figs.~\ref{fig:OD-1G-8o} and~~\ref{fig:IT-1G-8o} for offline gated
grant sizing, which does not limit the grant sizes, that the
improvements with optimized report message scheduling relative to
reporting at the beginning (and reporting at the beginning relative
to reporting at the end) become more pronounced for increasing
packet traffic load. With increasing traffic load, the individual
upstream transmission windows become longer, resulting in a larger
impact of the different report message scheduling approaches.

For online scheduling, we observe from
Figs.~\ref{fig:OD-1G-8o}--\ref{fig:IT-1G-32o} that the differences
between reporting at the end and reporting at the beginning
for $O = 8$ ONUs are magnified with respect to the
$O = 32$ case in Section~\ref{perf:sec} and in
Figs.~\ref{fig:OD-1G-32o}--~\ref{fig:IT-1G-32o_20}.
For the $O = 8$ case, each individual ONU upstream transmission
constitutes a relatively larger portion of the cycle,
increasing the relative impact of report message scheduling at the
beginning vs. the end.

For a moderately large number of $O = 32$ ONUs, we observe from
Figs.~\ref{fig:OD-1G-32o} and~\ref{fig:IT-1G-32o} that optimized
report message scheduling results in barely visible reductions of
packet delays and idle times only for offline gated grant sizing;
for the other offline DBA approaches there are no visible improvements. For
this moderately large number of ONUs, the individual upstream
transmission durations are relatively small compared to the overall
duration of a polling cycle, resulting in minimal potential for
improvements due to optimized report message scheduling.

As noted in Section~\ref{opt:sec}, for online scheduling, optimized
report message scheduling is not possible. Intuitively,
with online scheduling, the last ONU upstream transmission
$j = O$ in a cycle is not followed by a $2 \tau$ idle period.
Thus, there would be no benefit from reporting at the beginning
for the last ONU $j = O$.

For the 20~km range EPON, we observe from
Figs.~\ref{fig:OD-1G-32o_20} and~\ref{fig:IT-1G-32o_20} that for the
offline DBA approaches, similar to the observations for the
long-range EPON, reporting at the beginning reduces the mean packet
delays and idle times compared to reporting at the end. Optimized
report message scheduling reduces delays and idle times compared to
reporting at the beginning only very slightly for offline gated
grant sizing. For the short-range EPON, the mean idle times are
generally quite low, see Fig.~\ref{fig:IT-1G-32o_20}; thus there is
little potential to reduce idle times through optimized report
scheduling. Note in particular that for the 20~km EPON with a
round-trip propagation delay of approximately $2\tau = 0.2$~ms an
equal share of the maximum cycle duration, i.e., $Z/O$, can mask a
substantial portion of the $2 \tau$ for $Z = 2$ and 4~ms and all of
$2\tau$ for $Z = 8$~ms. We conclude therefore that for short-range
PONs, reporting at the beginning is an effective method for masking
the $2\tau$ idle time between successive cycles in offline
scheduling, especially for PONs with long cycle durations.

\bibliographystyle{IEEEtran}

\begin{thebibliography}{10}
\providecommand{\url}[1]{#1}
\csname url@samestyle\endcsname
\providecommand{\newblock}{\relax}
\providecommand{\bibinfo}[2]{#2}
\providecommand{\BIBentrySTDinterwordspacing}{\spaceskip=0pt\relax}
\providecommand{\BIBentryALTinterwordstretchfactor}{4}
\providecommand{\BIBentryALTinterwordspacing}{\spaceskip=\fontdimen2\font plus
\BIBentryALTinterwordstretchfactor\fontdimen3\font minus
  \fontdimen4\font\relax}
\providecommand{\BIBforeignlanguage}[2]{{%
\expandafter\ifx\csname l@#1\endcsname\relax
\typeout{** WARNING: IEEEtran.bst: No hyphenation pattern has been}%
\typeout{** loaded for the language `#1'. Using the pattern for}%
\typeout{** the default language instead.}%
\else
\language=\csname l@#1\endcsname
\fi
#2}}
\providecommand{\BIBdecl}{\relax}
\BIBdecl

\bibitem{MeMR14}
A.~Mercian, M.~McGarry, and M.~Reisslein, ``Impact of report message scheduling
  {(RMS)} in {1G/10G} {EPON} and {GPON},'' \emph{Optical Switching and
  Networking, in print}, 2014.

\bibitem{AuSMR10}
F.~Aurzada, M.~Scheutzow, M.~Maier, and M.~Reisslein, ``Towards a fundamental
  understanding of the stability and delay of offline {WDM EPONs},''
  \emph{IEEE/OSA Journal of Optical Communications and Networking}, vol.~2,
  no.~1, pp. 51--66, Jan. 2010.

\bibitem{ChAKLS10}
C.-H. Chang, N.~Alvarez, P.~Kourtessis, R.~Lorenzo, and J.~Senior,
  ``Full-service {MAC} protocol for metro-reach {GPONs},'' \emph{IEEE/OSA
  Journal of Lightwave Technology}, vol.~28, no.~7, pp. 1016--1022, Apr. 2010.

\bibitem{ChWCF11}
J.~Chen, L.~Wosinska, M.~N. Chughtai, and M.~Forzati, ``Scalable passive
  optical network architecture for reliable service delivery,'' \emph{IEEE/OSA
  J. Opt. Commun. and Netw.}, vol.~3, no.~9, pp. 667--673, Sep. 2011.

\bibitem{Dixit2013}
A.~Dixit, B.~Lannoo, G.~Das, D.~Colle, M.~Pickavet, and P.~Demeester,
  ``Flexible {TDMA/WDMA} passive optical network: Energy efficient
  next-generation optical access solution,'' \emph{Optical Switching and
  Networking}, vol.~10, no.~4, pp. 491--506, 2013.

\bibitem{EfE09}
F.~Effenberger and T.~El-Bawab, ``Passive optical networks {(PONs)}: Past,
  present, and future,'' \emph{Opt. Switch. Netw.}, vol.~6, no.~3, pp.
  143--150, 2009.

\bibitem{KeLE10}
K.~Kerpez, Y.~Luo, and F.~Effenberger, ``Bandwidth reduction via localized
  peer-to-peer {(P2P)} video,'' \emph{Int. J. Dig. Multim. Broadc.}, 2010.

\bibitem{KrDR12}
G.~Kramer, M.~DeAndrade, R.~Roy, and P.~Chowdhury, ``Evolution of optical
  access networks: Architectures and capacity upgrades,'' \emph{Proceedings of
  the IEEE}, vol. 100, no.~5, pp. 1188--1196, 2012.

\bibitem{MaMCW13}
M.~Mahloo, C.~Machuca, J.~Chen, and L.~Wosinska, ``Protection cost evaluation
  of {WDM}-based next generation optical access networks,'' \emph{Optical
  Switching and Networking}, vol.~10, no.~1, pp. 89--99, 2013.

\bibitem{SaS13}
G.~Sankaran and K.~Sivalingam, ``{ONU} buffer reduction for power efficiency in
  passive optical networks,'' \emph{Optical Switching and Networking}, vol.~10,
  no.~4, pp. 416--429, 2013.

\bibitem{WeAMR13}
X.~Wei, F.~Aurzada, M.~McGarry, and M.~Reisslein, ``{EIBT}: Exclusive intervals
  for bulk transfers on {EPONs},'' \emph{IEEE/OSA Journal of Lightwave
  Technology}, vol.~31, no.~1, pp. 99--110, Jan. 2013.

\bibitem{TanAH2010}
K.~Tanaka, A.~Agata, and Y.~Horiuchi, ``{IEEE} 802.3av 10{G-EPON}
  standardization and its research and development status,'' \emph{IEEE/OSA J.
  Lightw. Techn.}, vol.~28, no.~4, pp. 651--661, 2010.

\bibitem{XGPON}
``{ITU-T G.987.3}, \textit{10-Gigabit-capable passive optical networks
  (XG-PON): Transmission convergence (TC) specifications},''
  \url{http://www.itu.int/rec/T-REC-G.987.3/en}.

\bibitem{AnBT12}
M.~De~Andrade, A.~Buttaboni, M.~Tornatore, P.~Boffi, P.~Martelli, A.~Pattavina,
  and G.~Gavioli, ``Design of long-reach {TDM/WDM} passive optical access
  networks,'' in \emph{Proc. Telecommunications Network Strategy and Planning
  Symposium (NETWORKS)}, Oct. 2012, pp. 1--6.

\bibitem{Mou02}
A.~Helmy, H.~Fathallah, and H.~Mouftah, ``Interleaved polling versus
  multi-thread polling for bandwidth allocation in long-reach {PON}s,''
  \emph{IEEE/OSA J. Optical Commun. Netw.}, vol.~4, no.~3, pp. 210--218, Mar.
  2012.

\bibitem{Mou05}
B.~Kantarci and H.~Mouftah, ``Bandwidth distribution solutions for performance
  enhancement in long-reach passive optical networks,'' \emph{IEEE Commun.
  Surv. Tut.}, vol.~14, no.~3, pp. 714--733, Aug. 2012.

\bibitem{LiG13}
Y.~Liu, L.~Guo, C.~Yu, Y.~Yu, and X.~Wang, ``Planning of survivable long-reach
  passive optical network {(LR-PON)} against single shared-risk link group
  {(SRLG)} failure,'' \emph{Opt. Switching and Netw., in print}, 2013.

\bibitem{ScBLP11}
B.~Schrenk, F.~Bonada, J.~A. Lazaro, and J.~Prat, ``Remotely pumped long-reach
  hybrid {PON} with wavelength reuse in {RSOA-Based ONUs},'' \emph{IEEE/OSA J.
  Lightwave Techn.}, vol.~29, no.~5, pp. 635--641, Mar. 2011.

\bibitem{ShND13}
L.~Shi, A.~Nag, D.~Datta, and B.~Mukherjee, ``New concept in long-reach {PON}
  planning: {BER}-aware wavelength allocation,'' \emph{Optical Switching and
  Networking,}, vol.~10, no.~4, pp. 475--480, 2013.

\bibitem{SiSS12}
A.~Sivakumar, G.~C. Sankaran, and K.~M. Sivalingam, ``A comparative study of
  dynamic bandwidth allocation algorithms for long reach passive optical
  networks,'' \emph{IETE Techn. Rev.}, vol.~29, no.~5, pp. 405--413, 2012.

\bibitem{SKM0110}
H.~Song, B.~W. Kim, and B.~Mukherjee, ``Long-reach optical access networks: A
  survey of research challenges, demonstrations, and bandwidth assignment
  mechanisms,'' \emph{IEEE Communications Surveys and Tutorials}, vol.~12,
  no.~1, pp. 112--123, 1st Quarter 2010.

\bibitem{HoTr12}
D.~Hood and E.~Trojer, \emph{Gigabit-capable {PONs}}.\hskip 1em plus 0.5em
  minus 0.4em\relax Wiley, 2012.

\bibitem{GuPT12}
A.~Gumaste, K.~Pulverer, A.~Teixeira, J.~S. Wey, A.~Nouroozifar, C.~Badstieber,
  and H.~Schink, ``Medium access control for the next-generation passive
  optical networks: the {OLIMAC} approach,'' \emph{IEEE Network}, vol.~26,
  no.~2, pp. 49--56, March-April 2012.

\bibitem{AuSR11}
F.~Aurzada, M.~Scheutzow, M.~Reisslein, N.~Ghazisaidi, and M.~Maier, ``Capacity
  and delay analysis of next-generation passive optical networks {(NG-PONs)},''
  \emph{IEEE Trans. Commun.}, vol.~59, no.~5, pp. 1378--1388, May 2011.

\bibitem{DeSG10}
S.~De, V.~Singh, H.~Gupta, N.~Saxena, and A.~Roy, ``A new predictive dynamic
  priority scheduling in ethernet passive optical networks {(EPONs)},''
  \emph{Opt. Switching and Netw.}, vol.~7, no.~4, pp. 215--223, 2010.

\bibitem{KMP0202}
G.~Kramer, B.~Mukherjee, and G.~Pesavento, ``{IPACT}: A dynamic protocol for an
  {Ethernet PON (EPON)},'' \emph{IEEE Communications Magazine}, vol.~40, no.~2,
  pp. 74--80, February 2002.

\bibitem{LiRo13}
X.~Liu and G.~Rouskas, ``{MPCP}-$\ell$: Look-ahead enhanced {MPCP} for
  {EPON},'' in \emph{Proc. of IEEE ICC}, 2013.

\bibitem{LA0205}
Y.~Luo and N.~Ansari, ``{Bandwidth allocation for multiservice access on
  EPONs},'' \emph{IEEE Commun. Mag.}, vol.~43, no.~2, pp. S16--S21, Feb. 2005.

\bibitem{RaR11}
A.~Razmkhah and A.~G. Rahbar, ``{OSLG}: A new granting scheme in {WDM} ethernet
  passive optical networks,'' \emph{Optical Fiber Technology}, vol.~17, no.~6,
  pp. 586--593, Dec. 2011.

\bibitem{JiMFD12}
T.~Jimenez, N.~Merayo, P.~Fernandez, R.~Duran, I.~de~Miguel, R.~Lorenzo, and
  E.~Abril, ``Implementation of a {PID} controller for the bandwidth assignment
  in long-reach {PONs},'' \emph{IEEE/OSA J. Optical Commun. and Netw.}, vol.~4,
  no.~5, pp. 392--401, May 2012.

\bibitem{KT0809}
K.~Kanonakis and I.~Tomkos, ``Offset-based scheduling with flexible intervals
  for evolving {GPON} networks,'' \emph{IEEE/OSA Journal of Lightwave
  Technology}, vol.~27, no.~15, pp. 3259--3268, Aug. 2009.

\bibitem{QiXZ13}
Y.~Qin, D.~Xue, L.~Zhao, C.~K. Siew, and H.~He, ``A novel approach for
  supporting deterministic quality-of-service in {WDM} {EPON} networks,''
  \emph{Optical Switching and Networking}, vol.~10, no.~4, pp. 378--392, 2013.

\bibitem{BuAT13}
A.~Buttaboni, M.~De~Andrade, and M.~Tornatore, ``A multi-threaded dynamic
  bandwidth and wavelength allocation scheme with void filling for long reach
  {WDM/TDM PONs},'' \emph{IEEE/OSA Journal of Lightwave Technology}, vol.~31,
  no.~8, pp. 1149--1157, Apr. 2013.

\bibitem{Bur03}
B.~Kantarci and H.~Mouftah, ``{Periodic GATE Optimization (PGO): A New Service
  Scheme for Long-Reach Passive Optical Networks},'' \emph{IEEE Systems
  Journal}, vol.~4, no.~4, pp. 440--448, Dec. 2010.

\bibitem{MeMR13}
A.~Mercian, M.~McGarry, and M.~Reisslein, ``Offline and online multi-thread
  polling in long-reach {PONs}: A critical evaluation,'' \emph{IEEE/OSA J.
  Lightwave Technology}, vol.~31, no.~12, pp. 2018--2228, Jun. 2013.

\bibitem{Muk01}
H.~Song, B.-W. Kim, and B.~Mukherjee, ``Multi-thread polling: a dynamic
  bandwidth distribution scheme in long-reach {PON},'' \emph{IEEE J. Sel. Areas
  Commun.}, vol.~27, no.~2, pp. 134--142, Feb. 2009.

\bibitem{ZhMo09}
J.~Zheng and H.~Mouftah, ``A survey of dynamic bandwidth allocation algorithms
  for {Ethernet Passive Optical Networks},'' \emph{Optical Switching and
  Networking}, vol.~6, no.~3, pp. 151--162, Jul. 2009.

\bibitem{AuSH08}
F.~Aurzada, M.~Scheutzow, M.~Herzog, M.~Maier, and M.~Reisslein, ``Delay
  analysis of {Ethernet} passive optical networks with gated service,''
  \emph{OSA Journal of Optical Networking}, vol.~7, no.~1, pp. 25--41, Jan.
  2008.

\bibitem{DixDLCPD2011}
A.~Dixit, G.~Das, B.~Lannoo, D.~Colle, M.~Pickavet, and P.~Demeester,
  ``Adaptive multi-gate polling with void filling for long-reach passive
  optical networks,'' in \emph{Proc. ICTON}, Jun. 2011, pp. 1--4.

\bibitem{Mar01}
M.~McGarry and M.~Reisslein, ``Investigation of the {DBA} algorithm design
  space for {EPON}s,'' \emph{IEEE/OSA J. Lightw. Techn.}, vol.~30, no.~14, pp.
  2271--2280, Jul. 2012.

\bibitem{Kramer05}
G.~Kramer, \emph{Ethernet Passive Optical Networks}.\hskip 1em plus 0.5em minus
  0.4em\relax McGraw-Hill, 2005.

\bibitem{GPON}
``{ITU-T G.984.3}, \textit{Gigabit-capable Passive Optical Networks (G-PON):
  Transmission convergence layer specification},''
  \url{http://www.itu.int/rec/T-REC-G.984.3/en}.

\bibitem{RoyKHS11}
R.~Roy, G.~Kramer, M.~Hajduczenia, and H.~Silva, ``Performance of {10G-EPON},''
  \emph{IEEE Comm. Mag.}, vol.~49, no.~11, pp. 78--85, Nov. 2011.

\bibitem{BegHR2011}
P.~Begovic, N.~Hadziahmetovic, and D.~Raca, ``10{G} {EPON} vs. {XG-PON}1
  efficiency,'' in \emph{Proc. ICUMT}, 2011, pp. 1--9.

\bibitem{HajSM2007}
M.~Hajduczenia, H.~J.~A. da~Silva, and P.~Monteiro, ``{10G EPON} development
  process,'' in \emph{Proc. ICTON}, vol.~1, 2007, pp. 276--282.

\bibitem{Tal01}
G.~Talli, C.~W. Chow, E.~K. MacHale, and P.~D. Townsend, ``High split ratio
  116~km reach hybrid {DWDM-TDM} 10~{Gb/s} {PON} employing {R-ONUs},'' in
  \emph{Proc. ECOC}, Sep. 2006, pp. 1--3.

\bibitem{SCAW09}
B.~Skubic, J.~Chen, J.~Ahmed, L.~Wosinska, and B.~Mukherjee, ``A comparison of
  dynamic bandwidth allocation for {EPON}, {GPON}, and next-generation {TDM
  PON},'' \emph{IEEE Communications Magazine}, vol.~47, no.~3, pp. S40--S48,
  Mar. 2009.

\bibitem{IK1209}
H.~Ikeda and K.~Kitayama, ``Dynamic bandwidth allocation with adaptive polling
  cycle for maximized {TCP} throughput in {10G-EPON},'' \emph{IEEE/OSA J.
  Lightwave Techn.}, vol.~27, no.~23, pp. 5508--5516, Dec. 2009.

\bibitem{GutGS2011}
L.~Gutierrez, P.~Garfias, and S.~Sallent, ``Flexible joint scheduling {DBA} to
  promote the fair coexistence in {1G} and {10G} {EPON}s,'' in \emph{Proc.
  ONDM}, 2011, pp. 1--6.

\bibitem{Han2013}
M.-S. Han, ``Simple and feasible dynamic bandwidth allocation for {XGPON},'' in
  \emph{Proc. ICACT}, 2013, pp. 341--344.

\bibitem{HanYL2013}
M.~Han, H.~Yoo, and D.~Kee, ``Development of efficient dynamic bandwidth
  allocation algorithm for {XGPON},'' \emph{ETRI Journal}, vol.~35, no.~1, pp.
  18--26, Feb. 2013.

\bibitem{kra01}
G.~Kramer, B.~Mukherjee, and G.~Pesavento, ``Interleaved polling with adaptive
  cycle time {(IPACT)}: A dynamic bandwidth distribution scheme in an optical
  access network,'' \emph{Photonic Netw. Commun.}, vol.~4, no.~1, pp. 89--107,
  Jan. 2002.

\bibitem{AYDA1103}
C.~Assi, Y.~Ye, S.~Dixit, and M.~Ali, ``Dynamic bandwidth allocation for
  {Quality-of-Service} over {Ethernet PONs},'' \emph{IEEE J. Selected Areas in
  Commun.}, vol.~21, no.~9, pp. 1467--1477, November 2003.

\bibitem{BAS0706}
X.~Bai, C.~Assi, and A.~Shami, ``{On the fairness of dynamic bandwidth
  allocation schemes in Ethernet passive optical networks},'' \emph{Computer
  Communications}, vol.~29, no.~11, pp. 2123--2135, Jul. 2006.

\bibitem{Sam01}
S.~Choi, S.~Lee, T.-J. Lee, M.~Chung, and H.~Choo, ``Double-phase polling
  algorithm based on partitioned {ONU} subgroups for high utilization in
  {EPONs},'' \emph{IEEE/OSA Journal of Optical Communications and Networking},
  vol.~1, no.~5, pp. 484--497, Oct. 2009.

\bibitem{LuAn05}
Y.~Luo and N.~Ansari, ``Limited sharing with traffic prediction for dynamic
  bandwidth allocation and {QoS} provioning over {EPONs},'' \emph{OSA Journal
  of Optical Networking}, vol.~4, no.~9, pp. 561--572, 2005.

\bibitem{ZM0708}
Y.~Zhu and M.~Ma, ``{IPACT} with grant estimation {(IPACT-GE)} scheme for
  {Ethernet} passive optical networks,'' \emph{IEEE/OSA Journal of Lightwave
  Technology}, vol.~26, no.~14, pp. 2055--2063, Jul. 2008.

\bibitem{Mar03}
M.~McGarry, M.~Reisslein, F.~Aurzada, and M.~Scheutzow, ``Shortest propagation
  delay {(SPD)} first scheduling for {EPONs} with heterogeneous propagation
  delays,'' \emph{IEEE Journal on Selected Areas in Communications}, vol.~28,
  no.~6, pp. 849--862, Aug. 2010.

\bibitem{Luo13}
Y.~Luo, ``Activities, drivers, and benefits of extending {PON} over other
  media,'' in \emph{Proc. National Fiber Optic Engineers Conference}, 2013.

\bibitem{AuLMR14}
F.~Aurzada, M.~Levesque, M.~Maier, and M.~Reisslein, ``{FiWi} access networks
  based on next-generation {PON} and gigabit-class {WLAN} technologies: A
  capacity and delay analysis,'' \emph{IEEE/ACM Trans. Netw., in print}, 2014.

\bibitem{Bur04}
B.~Kantarci and H.~Mouftah, ``Energy efficiency in the extended-reach
  fiber-wireless access networks,'' \emph{IEEE Network}, vol.~26, no.~2, pp.
  28--35, March-April 2012.

\bibitem{Kun01}
K.~Yang, S.~Ou, K.~Guild, and H.-H. Chen, ``Convergence of {Ethernet PON} and
  {IEEE 802.16} broadband access networks and its {QoS}-aware dynamic bandwidth
  allocation scheme,'' \emph{IEEE Journal on Selected Areas in Communications},
  vol.~27, no.~2, pp. 101--116, Feb. 2009.

\bibitem{MiTK12}
M.~Milosavljevic, M.~Thakur, P.~Kourtessis, J.~Mitchell, and J.~Senior,
  ``Demonstration of wireless backhauling over long-reach {PONs},''
  \emph{IEEE/OSA J. Lightwave Techn.}, vol.~30, no.~5, pp. 811--817, 2012.

\bibitem{RaWL12}
C.~Ranaweera, E.~Wong, C.~Lim, and A.~Nirmalathas, ``Next generation
  optical-wireless converged network architectures,'' \emph{IEEE Network},
  vol.~26, no.~2, pp. 22--27, March-April 2012.

\bibitem{BiBC13}
A.~Bianco, T.~Bonald, D.~Cuda, and R.-M. Indre, ``Cost, power consumption and
  performance evaluation of metro networks,'' \emph{IEEE/OSA J. Opt.\ Comm.
  Netw.}, vol.~5, no.~1, pp. 81--91, Jan. 2013.

\bibitem{MaR04}
M.~Maier and M.~Reisslein, ``{AWG}-based metro {WDM} networking,'' \emph{IEEE
  Commun. Mag.}, vol.~42, no.~11, pp. S19--S26, Nov. 2004.

\bibitem{MaRW03}
M.~Maier, M.~Reisslein, and A.~Wolisz, ``A hybrid {MAC} protocol for a metro
  {WDM} network using multiple free spectral ranges of an arrayed-waveguide
  grating,'' \emph{Computer Netw.}, vol.~41, no.~4, pp. 407--433, Mar. 2003.

\bibitem{ScMRW03}
M.~Scheutzow, M.~Maier, M.~Reisslein, and A.~Wolisz, ``Wavelength reuse for
  efficient packet-switched transport in an {AWG}-based metro {WDM} network,''
  \emph{IEEE/OSA J. Lightw. Techn.}, vol.~21, no.~6, pp. 1435--1455, Jun. 2003.

\bibitem{YaMRC03}
H.-S. Yang, M.~Maier, M.~Reisslein, and W.~M. Carlyle, ``A genetic
  algorithm-based methodology for optimizing multiservice convergence in a
  metro {WDM} network,'' \emph{IEEE/OSA Journal of Lightwave Technology},
  vol.~21, no.~5, pp. 1114--1133, May 2003.

\bibitem{YuCL10}
M.~Yuang, I.-F. Chao, and B.~Lo, ``{HOPSMAN}: An experimental optical
  packet-switched metro {WDM} ring network with high-performance medium access
  control,'' \emph{IEEE/OSA J. Opt. Commun. Netw.}, vol.~2, no.~2, pp. 91--101,
  Feb. 2010.

\end{thebibliography}


\end{document}